\documentclass[floatfix, 12pt]{revtex4}
\usepackage[dvips,dvipdfmx]{graphicx}
\usepackage{soul}
\usepackage{amsmath}
\usepackage{amssymb}
\usepackage{amsfonts}
\usepackage{bmpsize}
\usepackage{enumerate}
\usepackage[english]{babel}
\usepackage{hyperref}
\usepackage{color}
\usepackage[utf8]{inputenc}
\usepackage{graphicx}
\usepackage{caption}
\usepackage{subcaption}
\usepackage{float}
\usepackage[normalem]{ulem}
\usepackage{multirow}
\usepackage{xcolor}

\newcommand{\nc}{\newcommand}
\nc{\be}{\begin{eqnarray}}
\nc{\ee}{\end{eqnarray}}
\newcommand{\bea}{\begin{eqnarray}}
\newcommand{\eea}{\end{eqnarray}}

\newcommand{\eq}[1]{Eq.~\eqref{#1}}
\newcommand{\alphem}{\alpha_{\rm em}}

\def\beq{\begin{equation}}
\def\eeq{\end{equation}}
\def\beqa{\begin{eqnarray}}
\def\eeqa{\end{eqnarray}}

\newcommand\sw{{\sin}\theta_{\mathrm{W}}}
\newcommand\cw{{\cos}\theta_{\mathrm{W}}}

\newcommand\sbeta{{\sin}{\beta}}
\newcommand\cbeta{{\cos}{\beta}}

\setstcolor{red}

\begin{document}

\begin{flushright}
CQUeST-2024-0744
\end{flushright}

\title{Experimental measurements outlasting MSSM-30 spectra}
\author{S.~AbdusSalam$^1$, S.S.~Barzani$^{1,2}$, M.~Mohammadidoust$^1$, S.A.~Ojaghi$^1$, L. Velasco-Sevilla$^{3,4}$}
\affiliation{$^1 \,\,$Department of Physics, Shahid Beheshti University, Tehran 1983969411, Iran
  \\ $^2 \,\,$ Department of Physics, Antwerp University, Belgium\\ $^3\,\,$ Center for Quantum Spacetime, Sogang University, Seoul 121-742, South Korea \\ $^4\,\,$ Department of Physics, Sogang University, Seoul 121-742, South Korea}

\begin{abstract}
Benchmarks for beyond the Standard Model (BSM) searches are mostly constructed around particular features of interest related to the experiment under consideration without giving due address to the results from other experiments. In this article, we present a 30-parameter MSSM phenomenological framework and benchmark points relevant for the long-lived particles and Higgs boson decays to BSM particle studies. These are extracted via scans of the model parameter space compatible with low-energy constraints from electroweak physics, B-physics, the dipole moment of the electron, and cold dark matter relic density measurements. We present four benchmark points that are illustrative on how light values of supersymmetric particles can still be. Neutralinos and charginos can be lighter than W-boson mass in sharp contrast to the expectations by conclusions derived based on simplified supersymmetric models.
\end{abstract}

\maketitle


\section{Introduction}
The discovery of the Higgs boson \cite{ATLAS:2012yve, CMS:2012qbp} is widely regarded as the final missing component of the standard model (SM) of particle physics. This can also be considered a success for the R-parity conserving minimal supersymmetric standard model (MSSM) \cite{Nilles:1983ge,Haber:1984rc,Gunion:1984yn} given the approximately prior-independent prediction for the Higgs boson mass based on the phenomenological MSSM global fits~\cite{AbdusSalam:2009qd, AbdusSalamThesis2009} just before the commissioning of the Large Hadron Collider (LHC).

Since there is no evidence for supersymmetric particles at the LHC, is this duly translated into a careful mapping of the MSSM parameters space, robustly delineating surviving versus ruled out regions?
  This is an important task with respect to the searches for BSMs. It is also an interesting avenue for testing current computing technologies and capabilities given the huge data from the LHC and non-collider experiments on the one hand, and the cursingly large number of free parameters from the theory side.

From the ATLAS and CMS collaborations, lower bounds on the possible mass scales for various SUSY particles are reported. These limits are, as expected and appreciated, strongly dependent on the benchmarks scenarios or the simplified representation of the supersymmetry(SUSY) models used for interpreting the LHC data. See for example the benchmarks presented in \cite{1109.3859, Bagnaschi:2021jaj}. Out of the various theoretical and phenomenological frames used for interpreting the LHC data, models with few number of parameters or selected sparticles, respectively like the CMSSM and simplified model scenarios are tightly constrained with bounds on the sparticle masses pushed well into the TeV or multi-TeV region for strongly interacting sparticles (see e.g.~\cite{CMS:2019ybf, CMS:2021cox}). But these and similar frames are not a complete representation for the MSSM as far as the SUSY-breaking parameter space coverage is concerned. The coverage or representation of the MSSM drastically increases when the pMSSM, MSSM-30, etc are considered in lieu of the CMSSM. 

From another perspective, with purely collider physics motivations, most of the MSSM benchmarks constructed are not necessarily compatible with the limits from low-energy particle physics, astro-particle, or cosmological experiments or observations. However, these limits can be used as guiding factors for mapping BSM hyper spaces.

We seek to pursue an approach that aims at complementing the traditional methods by addressing the gaps mentioned above. Eventually, the target is to make an exhaustive exploration of the MSSM parameter space using all (or as near all as possible) relevant experimental or observational inputs as guide for mapping and extracting benchmarks. In particular, we focus on a version of the MSSM with thirty free parameters~\cite{AbdusSalam:2014uea} (that is commonly referred as MSSM-30) and require compatibility with combined constraints from searches for sparticles and extra-SM Higgs bosons at colliders, dark matter, as well as constraints from the limit on the electric dipole of the electron. In this way, we set to determine MSSM regions/points, which are in the first place allowed with respect to these limits, for interpreting LHC data and making foresight with regards to proposed future colliders.

This article is organised as follows. We briefly review the MSSM-30 parametrisation of SUSY breaking in section~\ref{MSSM-30}. In section~\ref{Experimental-constraint} the experimental constraints and limits used while exploring the parameter space will be addressed. These include results from the LHC searches for SUSY and additional Higgs bosons as implemented in \texttt{SUSY-AI}~\cite{1605.02797}, \texttt{SMODELS}~\cite{Kraml:2013mwa, Ambrogi:2017neo, Dutta:2018ioj, Ambrogi:2018ujg} and \texttt{HiggsBounds}~\cite{Bechtle:2008jh, Bechtle:2011sb, Bechtle:2013wla} packages. The SUSY spectra are determined from the  base MSSM-30 parameters using \texttt{SPheno}~\cite{hep-ph/0301101, 1104.1573}. \texttt{FeynHiggs}~\cite{Frank:2006yh, Degrassi:2002fi} and \texttt{SusHi}~\cite{Harlander:2012pb, Harlander:2016hcx} were used for analysing the Higgs sector. The results and discussions are presented in section~\ref{theresults}.

\section{The MSSM-30}\label{MSSM-30}
A full description of the supersymmetry-breaking sector of the MSSM requires more than one hundred parameters \cite{hep-ph/9709450, 2401.03827}. The task of exploring such a vast parameter space in order to confront it with experimental data poses a significant challenge. To address this challenge, various well-motivated  {phenomenological frameworks} have been developed. One particular strategy involves reducing the number of free MSSM parameters by conducting a comprehensive{ and systematic} survey of SUSY-breaking parameters {compatible with observations}, such as {the absence of tree-level flavour changing neutral currents}. {In \cite{AbdusSalam:2014uea}, the} Minimal Flavour Violation (MFV) \cite{Hall:1990ac, Buras:2000dm, DAmbrosio:2002vsn, Buras:2003jf, Colangelo:2008qp} hypothesis was used for constructing the MSSM framework with forty-two and thirty free parameters, namely, MSSM-42 and MSSM-30 respectively. This strategy can be ameliorated with modern approach high dimensional space explorations based on artificial intelligence techniques -- which is the ultimate goal of our strategy and eventually to be reported in another work.

The MFV reduction of the MSSM parameters is based on a methodology which sought for a maximal probe of the MSSM parameter space with minimal imposition of ad hoc relations or truncations of the free parameters. Flavour physics effects are systematically represented as perturbation involving some expansion parameter: the sine of the Cabibbo mixing angle. Then the small size of flavour-violating effects are staged in terms of approximate symmetries. The group of SM flavour symmetries when all Yukawa couplings are neglected, G, get broken by spurion fields which are proportional to the SM Yukawa couplings. In this way, the action when expressed in terms of regular and spurion fields will be G-invariant. Replacing the spurion fields by their vacuum expectation values whose values are inspired from the SM Yukawa couplings will break symmetries in such a way that it will build automatically the GIM cancellations once loop effects are included. 
As such, all the low-scale SUSY breaking couplings can be constructed as powers of SM Yukawa coupling matrices, $Y_{U,D,E}$, such that all flavour violations are solely derived from the CKM matrix.

{The soft SUSY breaking terms expanded in G-invariant spurion factors are given by}~\cite{Ellis:2007kb,Colangelo:2008qp,DAmbrosio:2002vsn, Buras:2003jf,Hall:1990ac,Smith:2009hj}: 
\bea \label{mfvpars}
& &(M^2_Q)_{ij} = M^2_Q \left[ \delta_{ij} + b_1 (Y_U^\dagger
  Y_U)_{ij} + b_2 (Y_D^\dagger Y_D)_{ij} + c_1\{ (Y_D^\dagger Y_D
  Y_U^\dagger Y_U)_{ij} + H.c.\} + \ldots\right], \nonumber \\ \nonumber
& &(M^2_U)_{ij} = M^2_U \left[ \delta_{ij} + b_3 (Y_U
  Y_U^\dagger)_{ij} + \ldots \right],  \\ \nonumber
& &(M^2_D)_{ij} = M^2_D \left[
  \delta_{ij} + [Y_D (b_6 + b_7 Y_U^\dagger Y_U) Y_D^\dagger]_{ij}
 + \ldots \right],\\
& &(M^2_L)_{ij} = M^2_L \left[ \delta_{ij} + b_{13} (Y_E^\dagger
  Y_E)_{ij} + \ldots \right], \\ \nonumber
& &(M^2_E)_{ij} = M^2_E \left[ \delta_{ij} +
  b_{14} (Y_E Y_E^\dagger)_{ij} + \ldots \right] \,,
\eea
and
\bea \label{mfvpars2}
& &(A^{'}_E)_{ij} = a_E \left[ \delta_{ij} + b_{15} (Y_E^\dagger
  Y_E)_{ij} + \ldots \right], \nonumber\\ 
& &(A^{'}_U)_{ij} = a_U \left[ \delta_{ij} + b_9 (Y_U^\dagger
  Y_U)_{ij} + b_{10} (Y_D^\dagger Y_D)_{ij} + \ldots \right], \\
& &(A^{'}_D)_{ij} = a_D \left[ \delta_{ij} + b_{11} (Y_U^\dagger
  Y_U)_{ij} + b_{12} (Y_D^\dagger Y_D)_{ij} + c_6 (Y_D^\dagger Y_D
  Y_U^\dagger Y_U)_{ij} + \ldots \right].\nonumber
\eea
{The series reduce to only a few terms when the Cayley-Hamilton identities for $3 \times 3$ matrices are used. In general, the coefficients, $b_i$ and $c_i$, can take values that would span many orders of magnitude. The MFV hypothesis requires that $b_i$ and $c_i$ be all of order unity, while all the small numbers suppressing flavour changes solely arise from the products of the Yukawa matrices. The trilinear scalar couplings similarly take the form $(A_{E,U,D})_{ij} = (A^{'}_{E,U,D} Y_{E,U,D})_{ij}$. Working to all orders in the small Yukawa couplings, the MFV MSSM parameters will be the same as for the original MSSM. But the parameters can be reduced in a systematic way by neglecting terms smaller than a chosen order in small mixing angles such as the Cabibbo angle, $\theta_{CB}$, as expansion parameter $\lambda = \sin\theta_{CB} \simeq 0.23$.}  

{After the reduction of Eq.(\ref{mfvpars}) and Eq.(\ref{mfvpars2}) to finite series, large pieces of the terms such as $(Y_U^\dagger Y_U)_{ij}^2$ and $(Y_U^\dagger Y_U)_{ij}$ are found to be proportional to $V_{3i}^*V_{3j}$, with $V$ the CKM matrix. This term is followed by the next relatively smaller terms proportional to $V_{2i}^*V_{2j}$. In this way, $V_{3i}^* V_{3j}$ and $V_{2i}^* V_{2j}$ can replace $(Y_U^\dagger Y_U)_{ij}^2$ and $(Y_U^\dagger Y_U)_{ij}$ as basis vectors with coefficients of order one and $y_c^2 \sim \lambda^8$ respectively. In the same manner, $\delta_{i3}^*\delta_{j3}$ and $\delta_{i2}^* \delta_{j2}$ can replace $(Y_D^\dagger Y_D)_{ij}$ and $(Y_D^\dagger Y_D)_{ij}^2$ as basis vectors with coefficients $y_b^2$ and $y_s^2$ respectively where $\delta_{ij}$ are the unit matrix elements in family space. With this approach, all possible multipliable structures lead to the complete basis vectors, all of order unity, with a closed algebra under multiplication:}
\begin{eqnarray}
  \label{xbasis}
  \begin{array}{cccc}
    X_1 = \delta_{3i} \delta_{3j}, & X_2 = \delta_{2i} \delta_{2j}, &
    X_3 = \delta_{3i} \delta_{2j}, & X_4 = \delta_{2i} \delta_{3j},
    \\
    X_5 = \delta_{3i} V_{3j}, & X_6 = \delta_{2i} V_{2j}, & X_7 =
    \delta_{3i} V_{2j}, &X_8 = \delta_{2i} V_{3j}, \\
    X_9 = V^*_{3i} \delta_{3j}, & X_{10} = V^*_{2i} \delta_{2j}, &
    X_{11} = V^*_{3i} \delta_{2j}, & X_{12} = V^*_{2i} \delta_{3j},
    \\
    X_{13} = V^*_{3i} V_{3j}, & X_{14} = V^*_{2i} V_{2j}, & X_{15} =
    V^*_{3i} V_{2j}, & X_{16} = V^*_{2i} V_{3j}.
  \end{array}
\end{eqnarray}
{Thus, the MFV parameters can be assigned an order in $\lambda$ such that once the accuracy ${\cal O}(\lambda^n)$ of calculations is chosen then the above prescription can be used to systematically discard terms within the SUSY-breaking parameters expansion expressed in the $X_i$s. Keeping only terms of order ${\cal O}(\lambda^4) \sim {\cal O}(10^{-3})$ establishes the number of parameters to keep up, giving the MSSM-30 parameters:
\begin{eqnarray}
\label{mfvpar30}
& \widetilde{M}_1=e^{\phi_1} M_1, \quad \widetilde{M}_2=e^{\phi_2} M_2, \quad M_3, \quad \widetilde{\mu}=e^{\phi_\mu} \mu, \quad M_A, \quad \tan \beta, \nonumber\\
& M_Q^2=\tilde{a}_1 \mathbf{1}+x_1 X_{13}+y_1 X_1, \nonumber\\
& M_U^2=\tilde{a}_2 \mathbf{1}+x_2 X_1, \nonumber\\
& M_D^2=\tilde{a}_3 \mathbf{1}+y_3 X_1, \nonumber\\
& M_L^2=\tilde{a}_6 \mathbf{1}+y_6 X_1, \nonumber\\
& M_E^2=\tilde{a}_7 \mathbf{1}+y_7 X_1,\nonumber \\
& A^\prime_E=\tilde{a}_8 X_1, \nonumber\\
& A^\prime_U=\tilde{a}_4 X_5+y_4 X_1,\nonumber \\
& A^\prime_D=\tilde{a}_5 X_1+y_5 X_5 .
\end{eqnarray}

To perform a comprehensive exploration of the MSSM-30 parameter space, $\theta$,
\begin{eqnarray} \label{pars}
  \theta \equiv \left\{  \quad \operatorname{Re}\left[\widetilde{M}_{1,2}, \tilde{\mu}\right], M_3, M_A, \tan \beta, \quad \operatorname{Im}\left[\widetilde{M}_{1,2}, \tilde{\mu}\right], \quad a_{1,2,3,6,7}, \quad \operatorname{Re}\left[\tilde{a}_{4,5,8}\right], \right. \nonumber\\
   \left.\operatorname{Im}\left[\tilde{a}_{4,5,8}\right], \quad x_{1,2}, \quad y_{1,3,6,7}, \quad \operatorname{Re}\left[\tilde{y}_{4,5}\right], \quad \operatorname{Im}\left[\tilde{y}_{4,5}\right] \quad \right\},
\end{eqnarray}
we assigned the following ranges with the searches for new particles at the LHC in mind. Gaugino mass parameters $\widetilde{M}_1$ and $\widetilde{M}_2$ are allowed in -4 to $4  \, \mathrm{ TeV}$. The range for gluino mass parameter, $M_3$, is $100  \, \mathrm{ GeV}$ to 4 TeV. The parameters $a_{1,2,3,6,7}$ are sampled from within $(100  \, \mathrm{ GeV})^2$ to $(4  \, \mathrm{ TeV})^2$. Further, $x_{1,2}, y_{1,3,6,7}$ are allowed in the range $-(4  \, \mathrm{ TeV})^2$ to $(4  \, \mathrm{ TeV})^2$. The trilinear coupling terms $\operatorname{Re}\left[\tilde{a}_{4,5,8}\right], \operatorname{Im}\left[\tilde{a}_{4,5,8}\right], \operatorname{Re}\left[\tilde{y}_{4,5}\right]$, and $\operatorname{Im}\left(\tilde{y}_{4,5}\right)$ are taken from within $-8  \, \mathrm{ TeV}$ to $8  \, \mathrm{ TeV}.$ The pseudoscalar Higgs mass parameter, $m_A$ is allowed within $100  \, \mathrm{ GeV}$ to $4  \, \mathrm{ TeV}$ while the Higgs doublets mixing term, both real and imaginary parts $(\operatorname{Re}[\tilde{\mu}], \operatorname{Im}[\tilde{\mu}])$ are in the range -4 to $4  \, \mathrm{ TeV}$. The ratio of the vacuum expectation values $\tan \beta=\left\langle H_2\right\rangle /\left\langle H_1\right\rangle$ are allowed to be between 2 and $60$. This set of parameters are varied but with fixed relevant SM parameters $m_t^{\text {pole }}=173.1  \, \mathrm{ GeV}, \quad \alpha_s\left(M_Z\right)=0.118, \quad G_F=1.16637 \times 10^{-5}  \, \mathrm{ GeV}^{-2}, \quad m_b\left(m_b\right)=4.18  \, \mathrm{ GeV}, \quad M_Z=91.1876  \, \mathrm{ GeV}, \mathrm{ and  } \, M_W=80.385 \, \mathrm{ GeV}.$

\section{Sample scan and constraints} \label{Experimental-constraint}
In this section, we briefly explain  the limits applied to the MSSM-30 sample generated via random scans of the free parameters uniformly within the specified ranges.  The SUSY particle spectra and low-energy physics observables are computed using \texttt{SPheno}~\cite{hep-ph/0301101, 1104.1573}.

 Constraints {from searches from new particles as implemented in the package}  \texttt{SMODELS} \cite{Kraml:2013mwa, Ambrogi:2017neo, Dutta:2018ioj, Ambrogi:2018ujg}, {from precision Higgs measurements in} \texttt{HiggsBounds}~\cite{Bechtle:2008jh, Bechtle:2011sb, Bechtle:2013wla}, from dark matter searches related limits with neutralino LSP as dark matter candidate as implemented in \texttt{micrOMEGAs}~\cite{hep-ph/0112278,hep-ph/0405253,Belanger:2008sj, 2003.08621, 1305.0237}, and {those from the electron's electric dipole moment (EDM)~\cite{ACME:2018yjb, Roussy:2022cmp} limits were imposed for selecting the sample points.

Only the parameter space points that pass all the constraints and limits from experiments imposed were accepted and saved for further processing. The model parameters were first passed to \texttt{SPheno}~\cite{hep-ph/0301101, 1104.1573} which generates the particles spectra, couplings and mixings. It also performs checks on the generated spectra to be acceptable with respect to certain physical and theoretical requirements such as the electroweak symmetry breaking conditions. The lightest neutralino is required to be the lightest supersymmetric particle (LSP). In addition, SUSY spectra were accepted if the Higgs boson mass falls within 122 to 128 GeV range. Here the $\pm 3$ GeV range relative to the experimentally measured value around 125 GeV is allowed as the ball-pack uncertainty in the predictions of the Higgs boson masses by \texttt{SPheno}. Using \texttt{micrOMEGAs}~\cite{hep-ph/0112278,hep-ph/0405253,Belanger:2008sj, 2003.08621, 1305.0237}, the dark matter direct detection search and LHC limits implemented in \texttt{HiggsBounds}~\cite{Bechtle:2008jh, Bechtle:2011sb, Bechtle:2013wla} were imposed. It turned out to be computationally extremely hard to find MSSM-30 parameters that simultaneously satisfy these set of requirements. For the analyses presented here, we work with a sample consisting of about 86000 points that passed the filterings stated so far~\footnote{This took most of the HTCondor cluster resources we had access to at CERN for a couple of weeks.}. We think that this sample should have already captured a diverse representative features of model space configurations and physics possibilities. In the sub-sections that follow, we address the categories of constraints for extracting correlations and features of the MSSM-30 sample.

First, sample points are drawn uniformly within the specified ranges of allowed values for the MSSM-30 parameters. These are then transformed into the SUSY breaking terms~Eq.(\ref{pars}) using the SUSY Les Houches Accord 2 (SLHA-2) format~\cite{0801.0045} which are subsequently passed to \texttt{SPheno} for generating sparticles spectra. Only parameter points with successfully generated spectra having the lightest neutralino as the LSP, and lightest Higgs boson mass within the range 123 to 128 GeV, allowing for $\pm 3$ GeV theoretical error,  are considered acceptable for further processing. Following this, the spectra are then assessed with respect to the collider limits implemented in \texttt{HiggsBounds} and the dark matter direct detection limits in \texttt{micrOMEGAs}. The MSSM-30 sample gathered in this way is then further post-processed using limits from searches for SUSY as implemented in \texttt{SModelS} and \texttt{SUSY-AI}, as well as the limit from the search for electron EDM. 

\paragraph{Higgs measurements and limits}
One of the fundamental aspects of research related to the LHC is directed towards ascertaining whether the observed 125 GeV state is part of a larger Higgs sector with multiple distinct new particles linked to electroweak symmetry breaking. To facilitate the utilisation of  information from Higgs searches and measurements at LEP, the Tevatron, and the LHC, particularly in terms of relatively model-independent cross-section limits for testing various theoretical models, the development of the \texttt{HiggsBounds} program \cite{2006.06007} has been instrumental. We apply the set of Higgs measurements-related set of constraints for mapping the MSSM-30 parameters space. Given the MSSM-30 Higgs sector masses, mixing angles and couplings, it uses the experimental topological cross section limits from Higgs searches at LEP, the Tevatron and the LHC to determine whether the MSSM point is excluded or not at 95\% confidence level. The software package \texttt{SUSYPOPE}~\cite{Heinemeyer:2006px,Heinemeyer:2007bw} were used for checking the compatibility of the Z-boson decay width predictions with the measured value~\cite{ParticleDataGroup:2024cfk}.

\paragraph{Limits from searches for SUSY particles}
For beyond the SM (BSM) with $Z_2$-like symmetry such as the MSSM, an approximate way for reusing or reinterpreting limits from the LHC searches for new particles is possible using the \texttt{SModelS} program~\cite{Kraml:2013mwa, Ambrogi:2017neo, Dutta:2018ioj, Ambrogi:2018ujg}. This is based on the simplified model approach but can be useful for probing MSSM spectra. MSSM spectra can be decomposed into various simplified model scenarios. We find that amongst the LHC limits implemented in \texttt{SModelS}, \cite{ATLAS:2019gqq, CMS:2016ybj} are the most constraining on the MSSM-30 sample analysed. \texttt{SModelS} classifies the MSSM-30 points as being ``allowed'', ``excluded'', or ``not tested'' based on whether $r \geq 1$ or not. Here $r$ is the ratio of the theoretical cross section prediction for a reduced model topology over the corresponding observed upper limit at the LHC as implemented in the software package. We also used the phenomenological MSSM and machine-learning based limits~\cite{1508.06608, 1605.09502} implemented in \texttt{SUSY-AI}~\cite{1605.02797}. Again, this is considered another approximate probe of the MSSM-30 since \texttt{SUSY-AI} requires MSSM spectra in SLHA-1 format, which can be derived from the full-fledged SLHA-2 spectra of the MSSM-30 sample. 

\paragraph{Limit from search for electron's EDM} 
The question related the dominance of matter over antimatter in our observed universe can possibly be addressed within the context of the MSSM-30. At high-energy or very small length scales, sparticles can manifest in the vacuum and interact with say an electron in a way that alters its characteristics. Time-reversal violating or CP-violating interactions will contribute to the  electron EDM, $d_e$. Therefore the limits from experiments searching for such phenomenon can be used to delineate the MSSM parameters space. With this, we address the interesting research direction whereby the interplay of limits from searches for electron EDM and limits from the searches for SUSY at colliders are seriously taken into consideration for analysing BSMs. Along this line of thought, MSSM-30 benchmark points which are compatible with the EDM limit~\cite{ACME:2018yjb, Roussy:2022cmp} will be presented. The SUSY spectra are allowed or excluded using the $90\%$ confidence level lower bound, $ |d_e| = 4.1 \times 10^{-30}\, \textrm{e} \, \textrm{cm}$. 

\paragraph{Dark matter related constraints}
The fact that no dark matter (DM) particle is detected so far is used to put limits on the neutralino-nucleons cross section based on some assumed local neighbourhood DM density and velocity distribution (see e.g. \cite{Schumann:2019eaa}). The spin-independent neutralino-nucleus elastic scattering cross section is given by  
\be
\sigma \approx \frac{4 m^2_{\tilde \chi^0_1} m^2_{T}}{\pi (m_{\tilde
    \chi^0_1} + m_T)^2} [Z f_p + (A-Z) f_n]^2,
\ee
where $m_T$ is the mass of the target nucleus and $Z$ and $A$ are the atomic number and atomic mass of the nucleus, respectively.  $f_p$ and $f_n$ are neutralino couplings to protons and neutrons, given by~\cite{Gelmini:1990je,Drees:1992rr,Drees:1993bu,Jungman:1995df,Ellis:2000ds} 
\be 
f_{p,n} = \sum_{q=u,d,s} f^{(p,n)}_{T_q} a_q \frac{m_{p,n}}{m_q} +
\frac{2}{27} f^{(p,n)}_{TG} \sum_{q=c,b,t} a_q  \frac{m_{p,n}}{m_q}, 
\label{quds}
\ee
where $a_q$ are the neutralino-quark couplings and $f^{(p,n)}_{T_q}$ represents the quark content of the nucleon. For the MSSM-30, only parameter space points that pass the limits from LUX \cite{Akerib:2015rjg}, PANDA \cite{Cui:2017nnn} and XENON \cite{Aprile:2018dbl} experiments, as implemented in \texttt{micrOMEGAs}, and whose DM thermal relics density do not over close the universe, were accepted and kept for further analyses.

\section{Results and discussions}   \label{theresults}
In this section, we address the effect of the various constraints considered on the MSSM-30 sample.

\paragraph{Limits from searches for SUSY} For the MSSM-30 base parameters, the change in the number of model points per bin, starting with total sample gathered from the random scans, due to requiring compatibility with LHC limits as implemented in \texttt{SModelS} and \texttt{SUSY-AI} are shown in Fig.~\ref{dparams1} and Fig.~\ref{dparams2}. Each of the plots shows the histogram of parameters comparing various sub-samples allowed by \texttt{SUSY-AI} (magenta colour) and by \texttt{SModelS} (orange). The scenario whereby the MSSM-30 point is not tested by \texttt{SModelS} but allowed by \texttt{SUSY-AI} is also shown (green). In the figures, only the parameters with distinct histogram shapes are shown. The $M_1$ and $M_2$ histograms are similar to their corresponding imaginary parts. Similarly, $a_1$ histogram is representative for $a_2, a_3, a_6, a_7$. The same is the case for $a_4$ to $Im(a_4), a_5, Im(a_5), y_4, Im(y_4), y_5, Im(y_5), x_1, a_8$ parameters; and $Im(a_8)$ to $y_1; y_3; x_2; y_6, y_7$ parameters.

According to the plots, the pseudoscalar Higgs and gluino mass parameters approximately have to be greater than $800$ GeV and $1.5$ TeV respectively. MSSM-30 gluino mass parameters beyond $2.5$ TeV are not probed by \texttt{SModelS}. About one-quarter of the sample points are excluded by \texttt{SModelS}. Again, this should be taken with care since the MSSM-30 spectra have complete SUSY states, so the simplified model results are only indicative but can be taken on-board analyses as we present in the article. More than one-third of the sample points are not probed and another one-third are allowed. 

\begin{figure}[!ht] 
  \includegraphics[angle=0, width=0.43\textwidth]{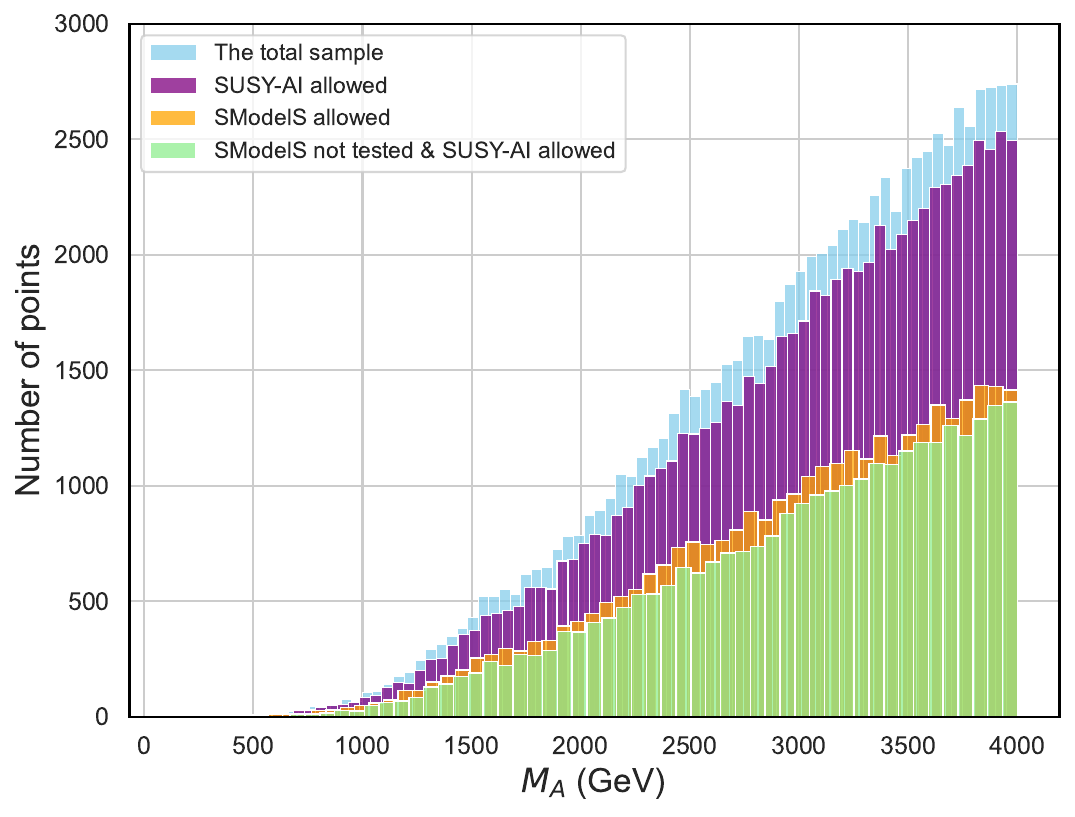}
  \includegraphics[angle=0, width=0.43\textwidth]{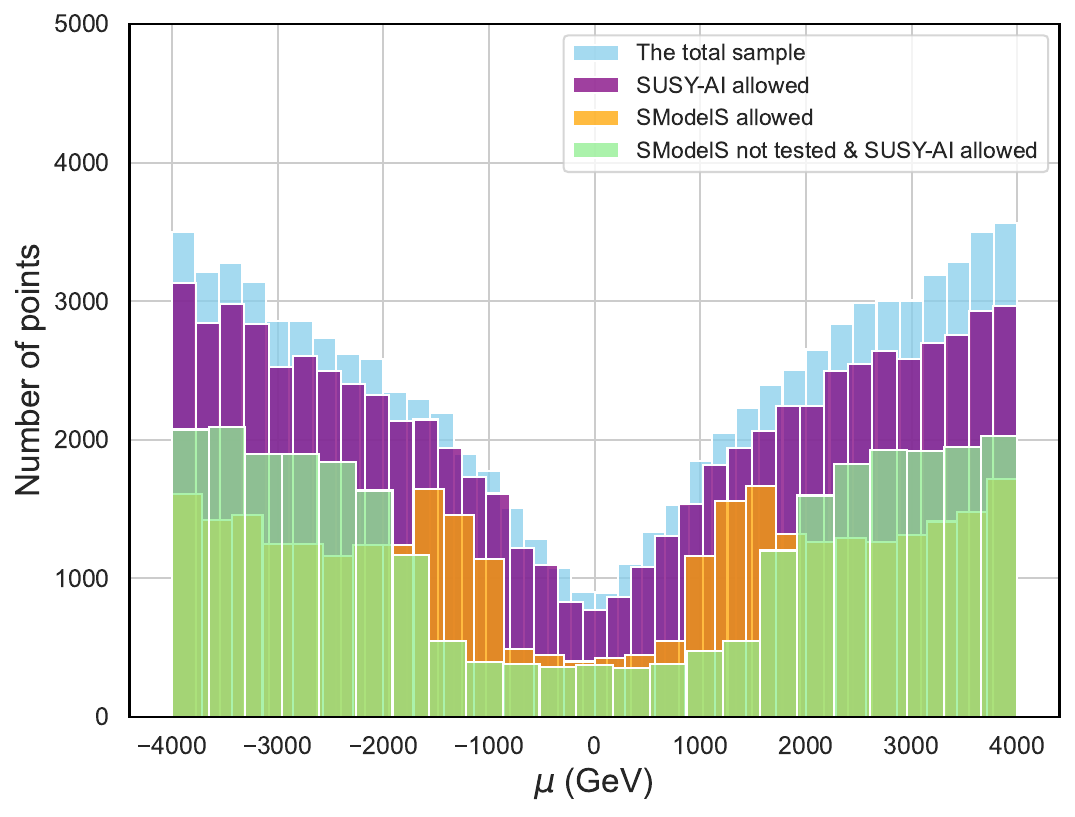}  \\
  \includegraphics[angle=0, width=0.43\textwidth]{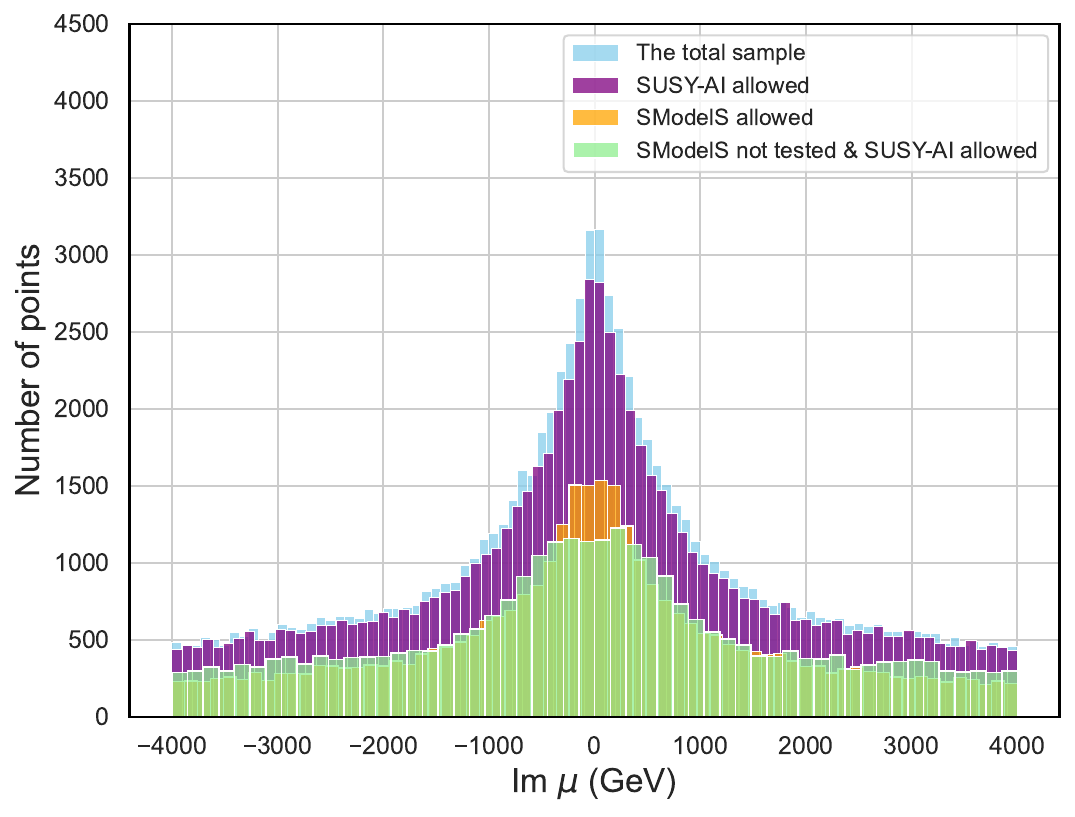}
  \includegraphics[angle=0, width=0.43\textwidth]{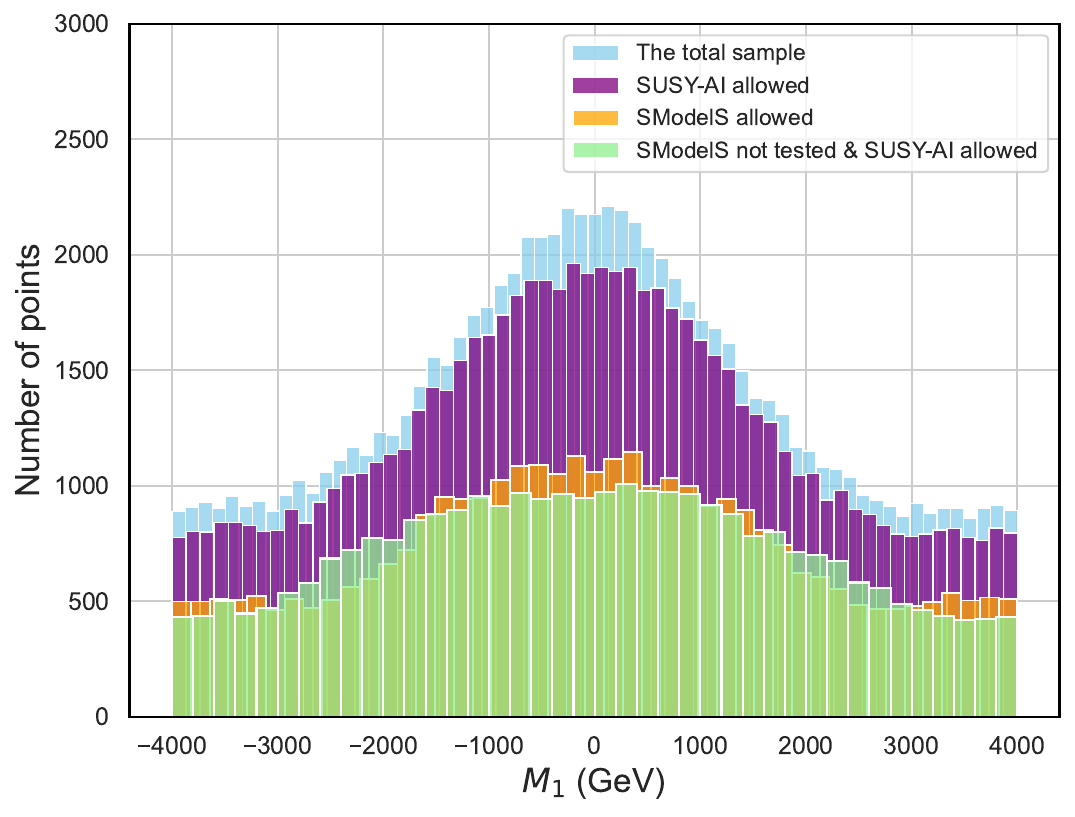}  \\
  \includegraphics[angle=0, width=0.43\textwidth]{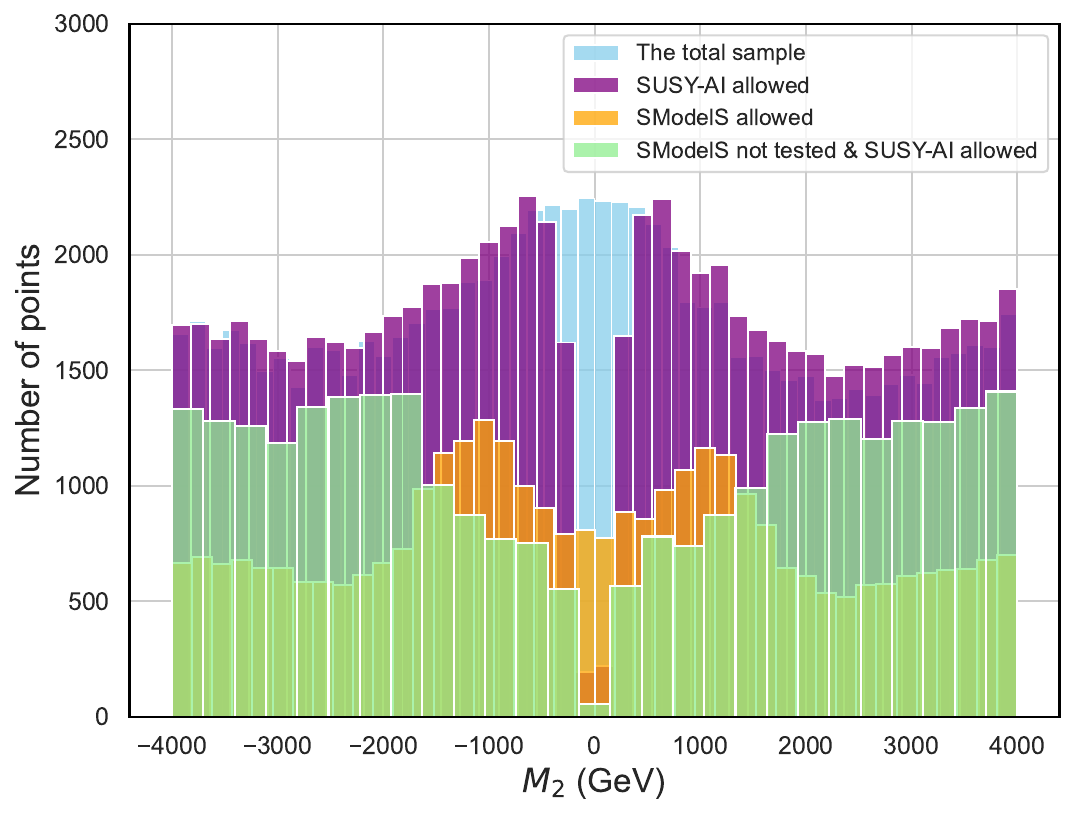}
  \includegraphics[angle=0, width=0.43\textwidth]{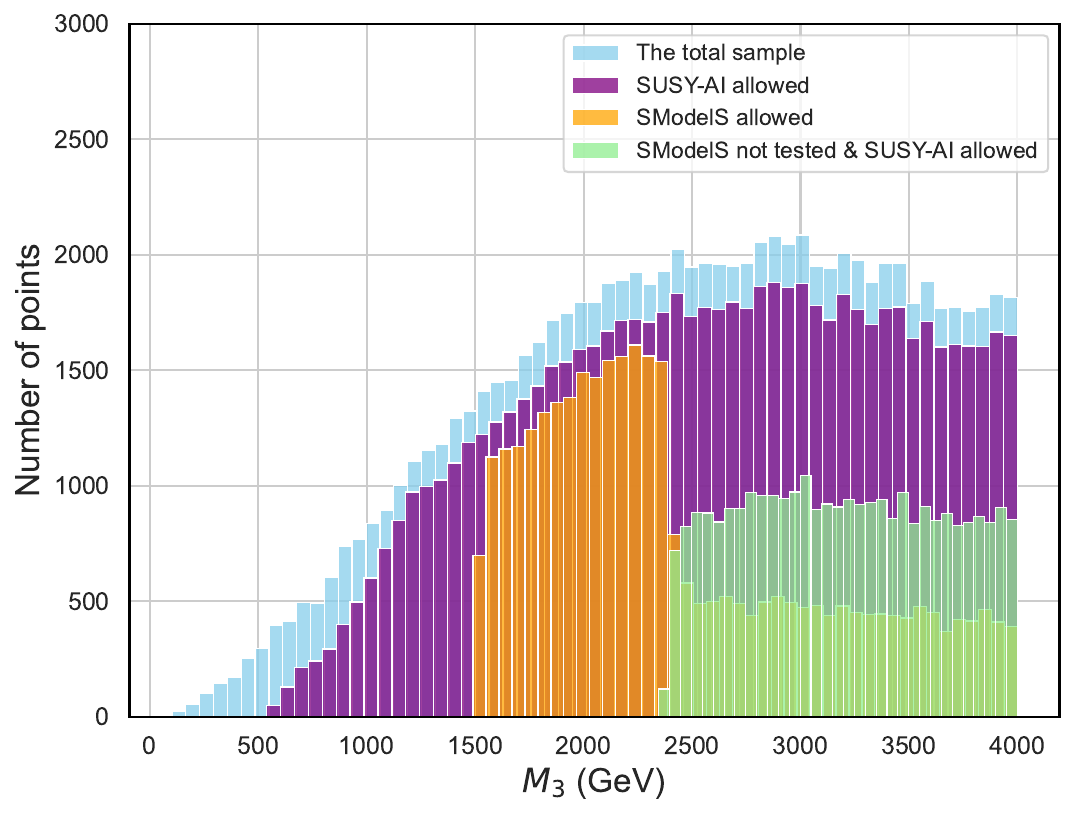}  \\
  \includegraphics[angle=0, width=0.43\textwidth]{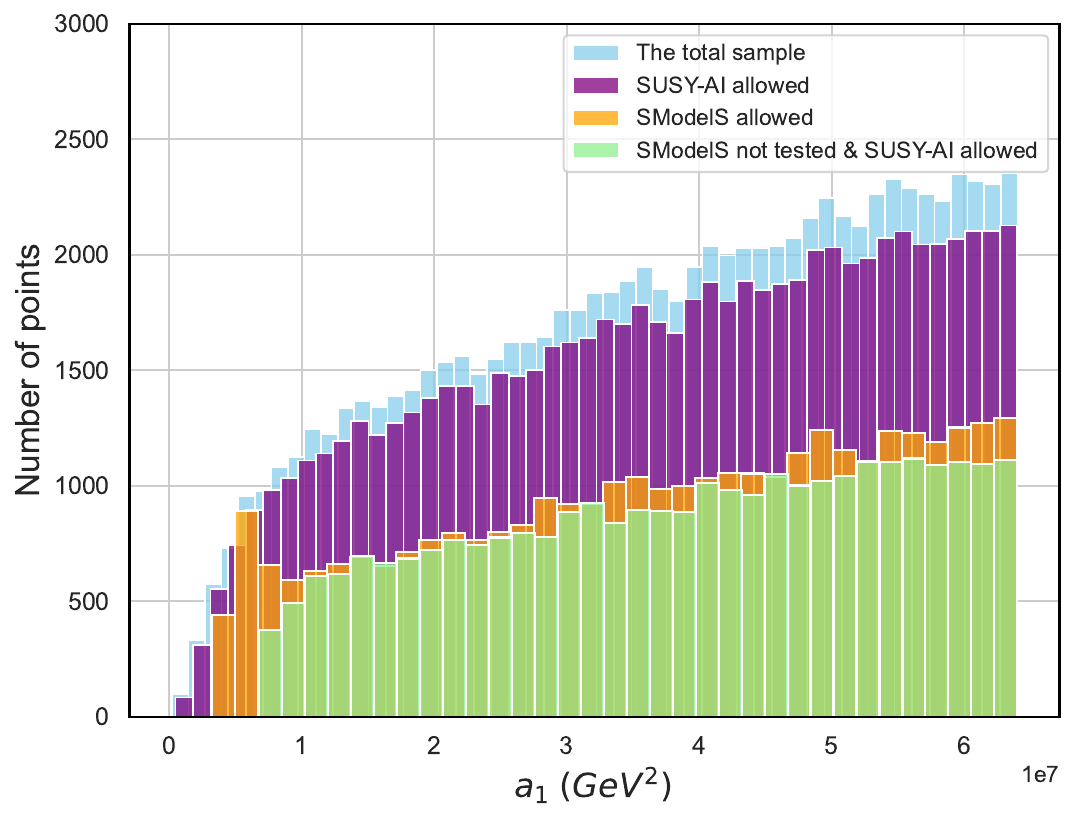}
  \includegraphics[angle=0, width=0.43\textwidth]{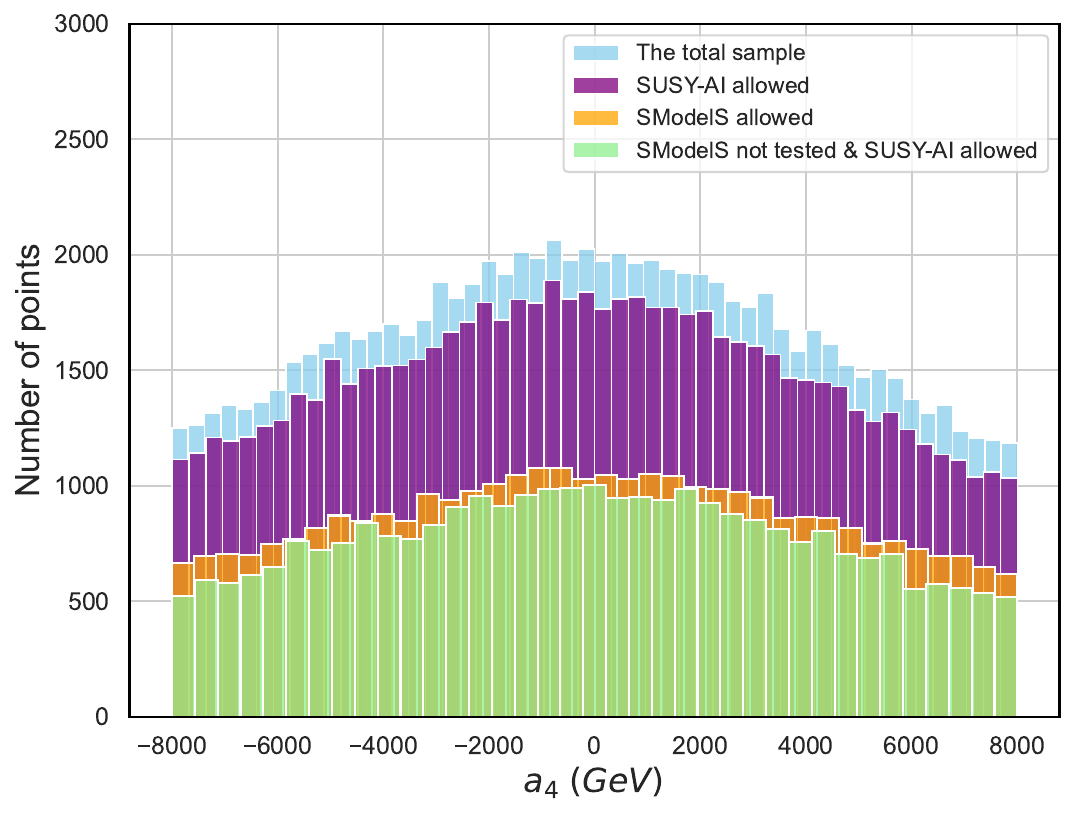}
  \caption{The histograms of MSSM-30 free parameters comparing the distributions for samples that survive SModelS and SUSY-AI constraints. From top to bottom: $M_A$, $\mu$, $Im(\mu)$, $M_1 \sim Im(M_1)$, $M_2 \sim Im(M_2)$, $M_3$, $a_1 \sim a_{2,3,6,7}$ and $a_4 \sim a_{5,8}$  $\sim Im(a_{4,5})$   $\sim y_{4,5}$   $\sim Im(y_{4,5})$  $ \sim x_1$. Here the symbol $\sim$ means that the distributions are similar.} 
  \label{dparams1}
\end{figure}

\begin{figure}[!ht] 
  \includegraphics[angle=0, width=0.43\textwidth]{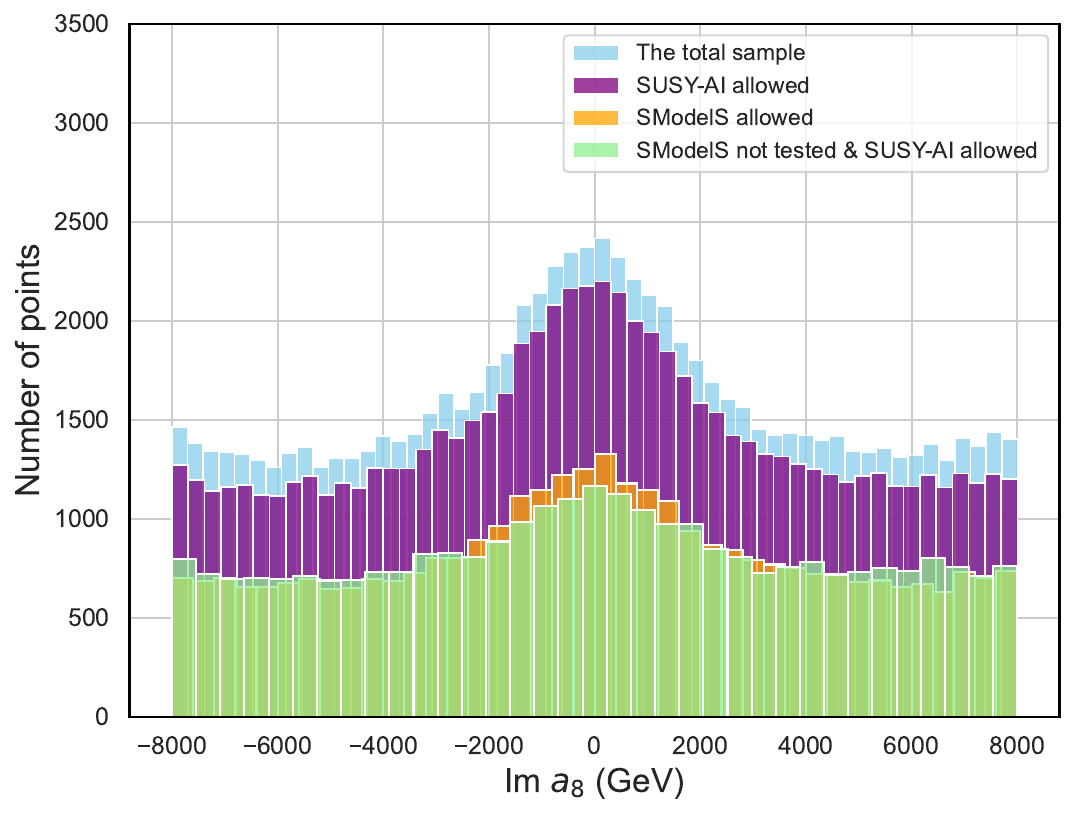}
  \includegraphics[angle=0, width=0.43\textwidth]{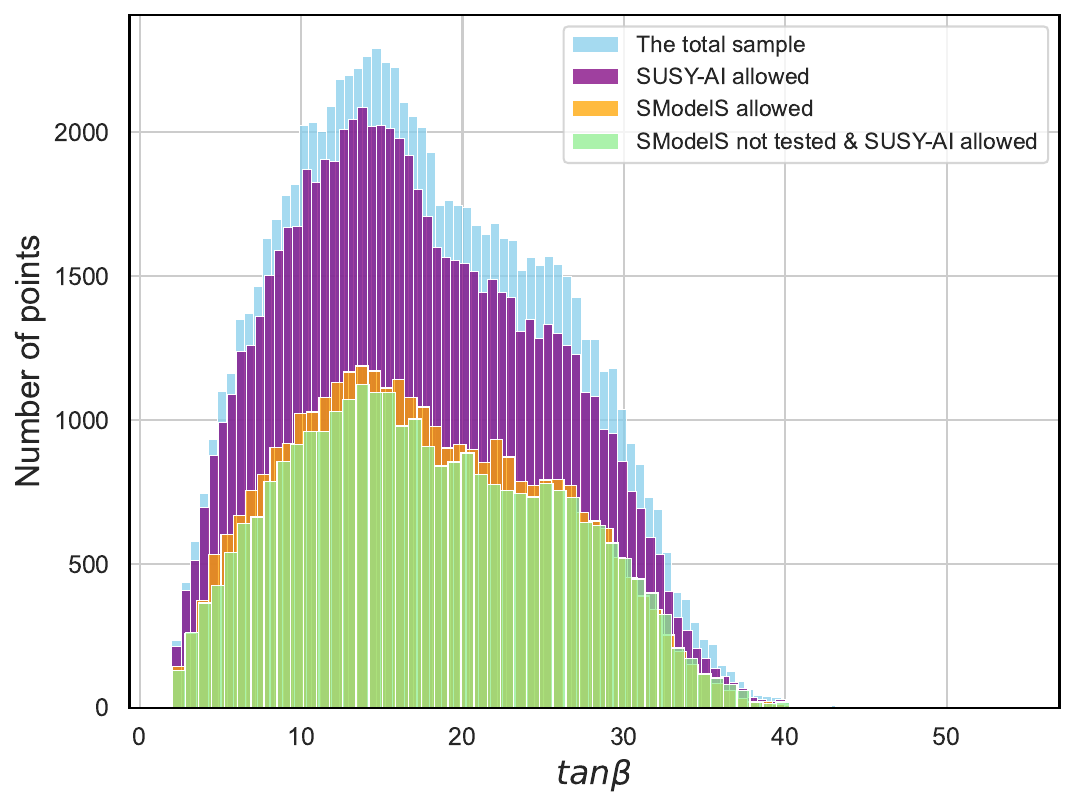}
   \caption{The histograms of MSSM-30 free parameters comparing the distributions for samples that survive SModelS and SUSY-AI constraints. From left to right: $Im(a_8) \sim y_{1,3,6,7} \sim x_2$ and $\tan\beta$. Here the symbol $\sim$ means that the distributions are similar.} 
  \label{dparams2}
\end{figure}

Similar to the base parameters, in Fig.~\ref{observs1} and Fig.~\ref{observs2} we show the impact of the limits from search for SUSY at the LHC on MSSM-30 predictions for the electron EDM, LSP dark matter candidate relic density and sparticles masses. \texttt{SModelS} and \texttt{SUSY-AI} limits do not effect the distribution of the first two observable. A wide range of neutralino LSP, even within the much less than 1 TeV region, is still unprobed. The same is the case for sub-TeV to order TeV lightest charginos are still viable.  A wide range of neutralino LSP, even within the much less than 1 TeV region, is still unprobed. The same is the case for order TeV lightest charginos. The SUSY limits made the most impact on the chargino and scalar charm masses. As before, only the parameters with distinct histogram shapes are shown. The $m_H$ histogram shape is similar to that for $m_H^\pm$. In the same way, the $\tilde{b}_1$ plot is also representing for $\tilde{u}_R$, $\tilde{d}_R$, $\tilde{e}_R$, and $\tilde{t}_1$. $\tilde{b}_2$ and $\tilde{t}_2$ masses turn out to be much greater than 4 TeV that one can claim they will be unaccessible at the LHC. The $\tilde{c}_L$ plot is also representing for $\tilde{d}_L$, $\tilde{s}_L$, and $\tilde{u}_L$; $\tilde{c}_R$ for $\tilde{s}_R$, and $\tilde{\tau}_2$; $\tilde{\mu}_R$ for $\tilde{\nu}_{\tau L}$; and $\tilde{\nu}_{e L}$ for $\tilde{\nu}_{\mu L}$.

\begin{figure}[!ht] 
  \includegraphics[angle=0, width=0.43\textwidth]{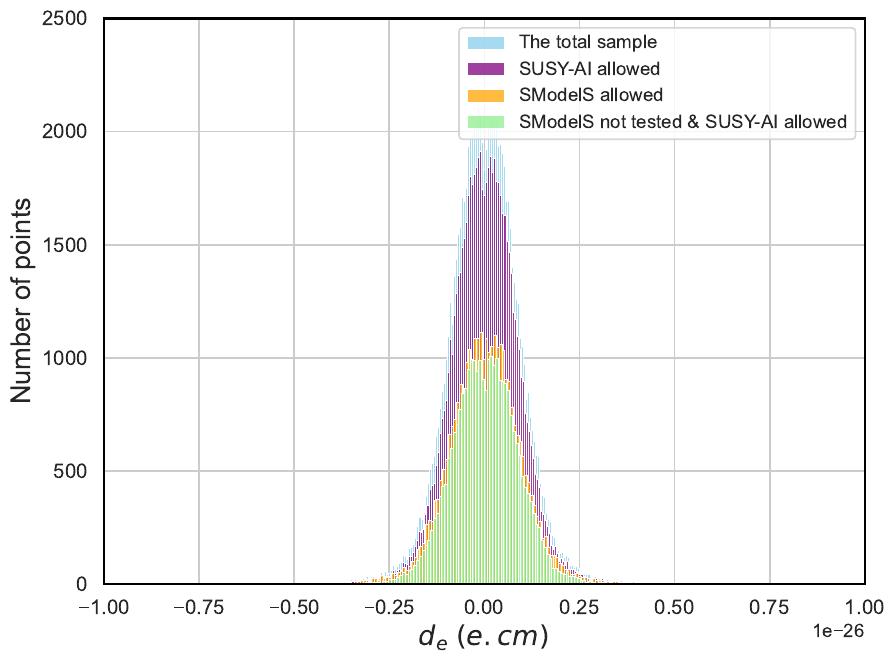}
 \includegraphics[angle=0, width=0.43\textwidth]{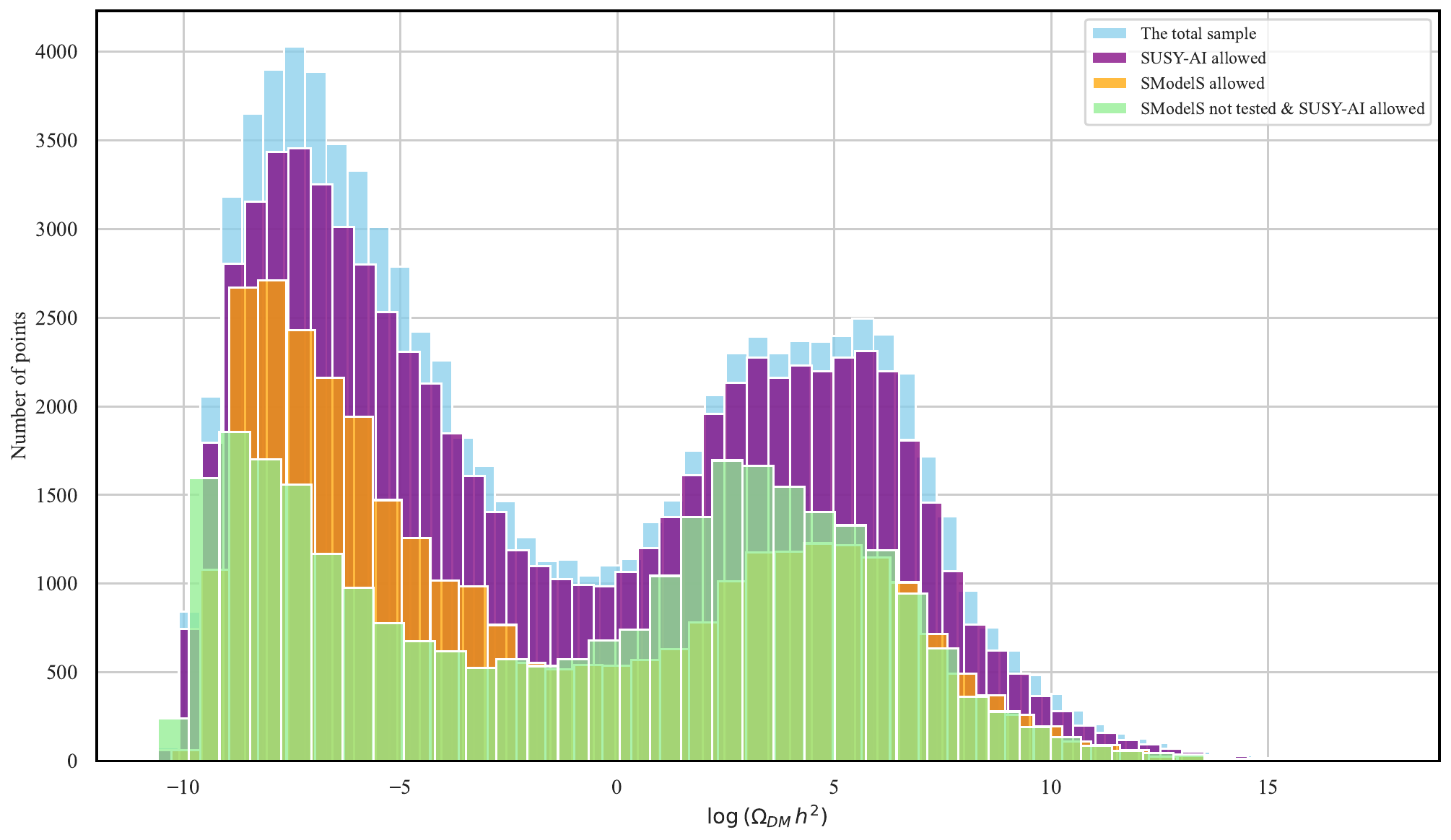} \\  
 \includegraphics[angle=0, width=0.43\textwidth]{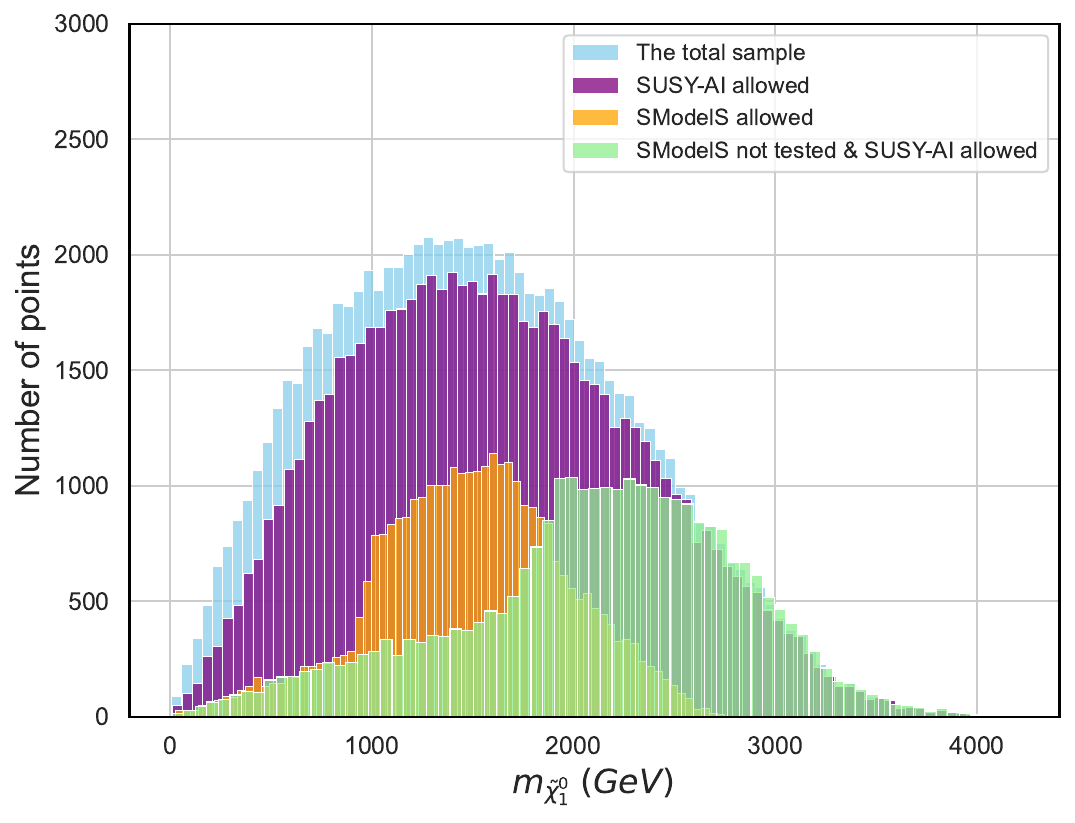}
 \includegraphics[angle=0, width=0.43\textwidth]{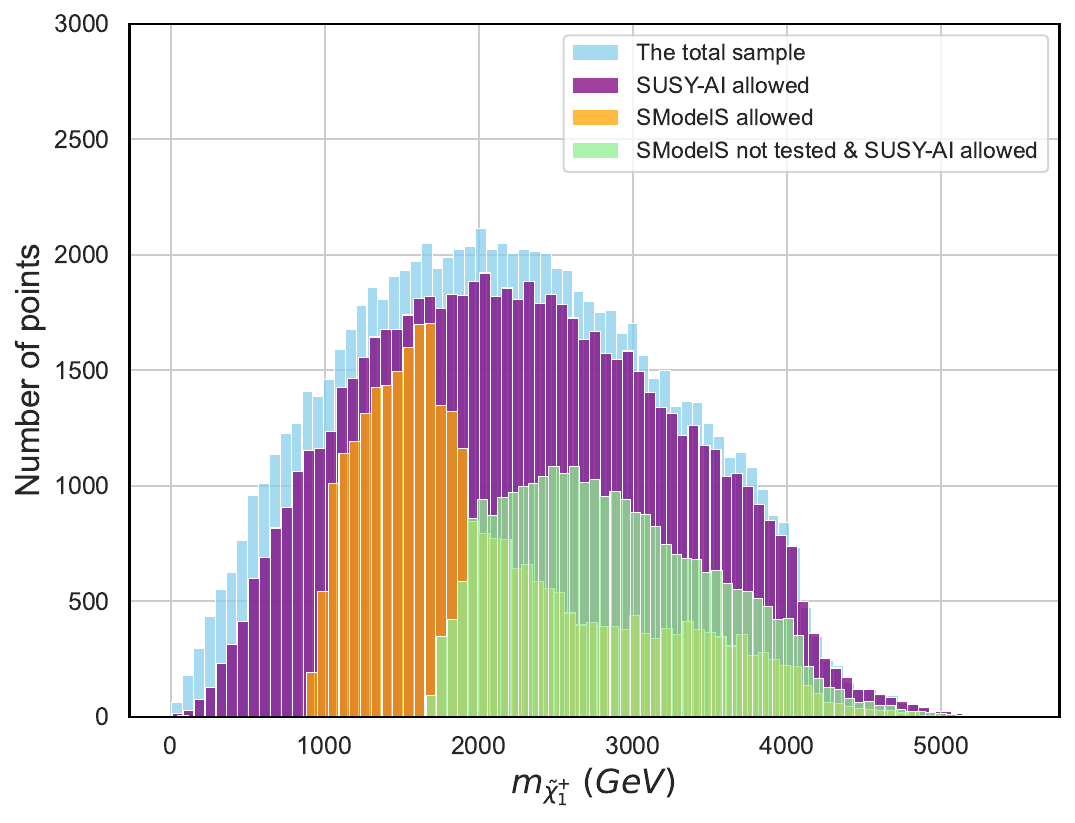} 
  \includegraphics[angle=0, width=0.43\textwidth]{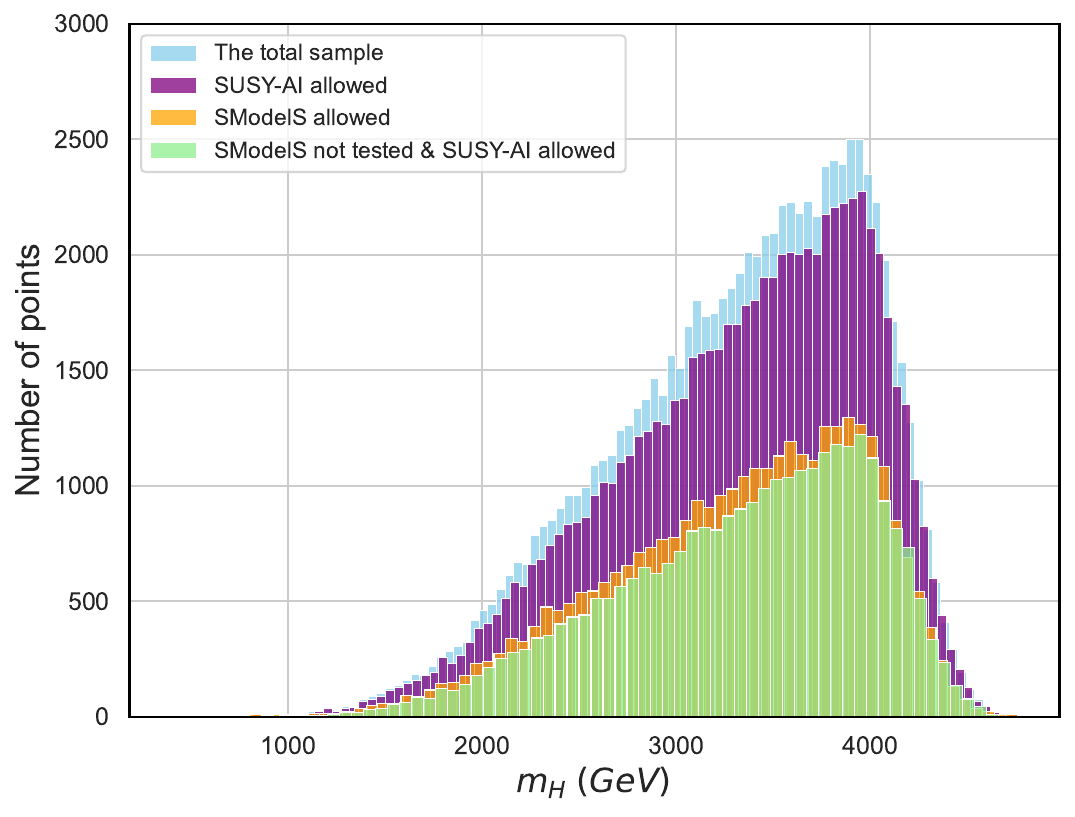} 
 \includegraphics[angle=0, width=0.43\textwidth]{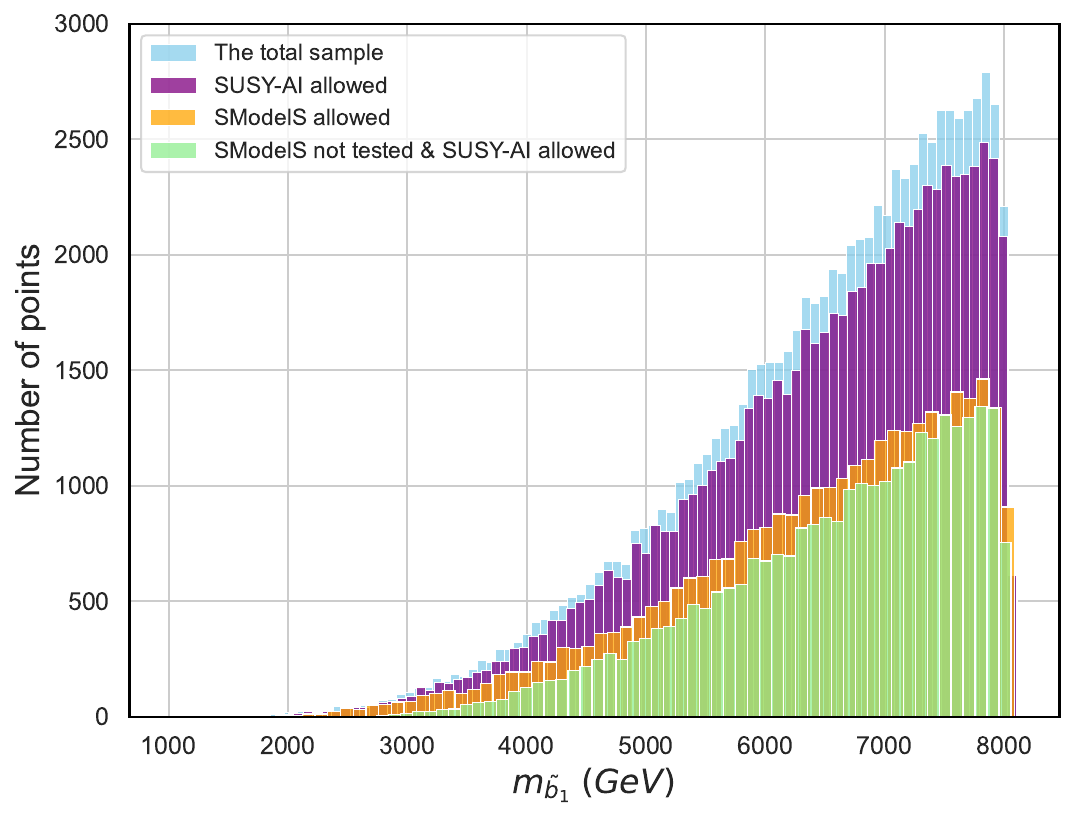}\\
 \includegraphics[angle=0, width=0.43\textwidth]{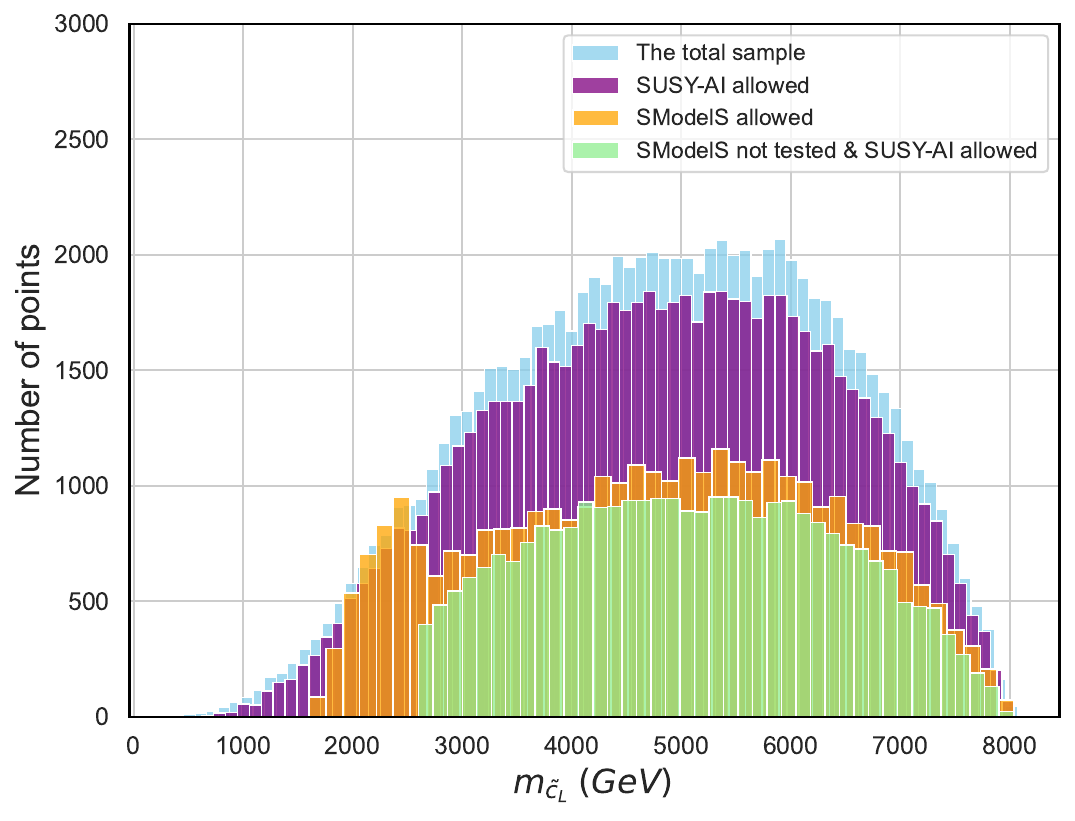} 
 \includegraphics[angle=0, width=0.43\textwidth]{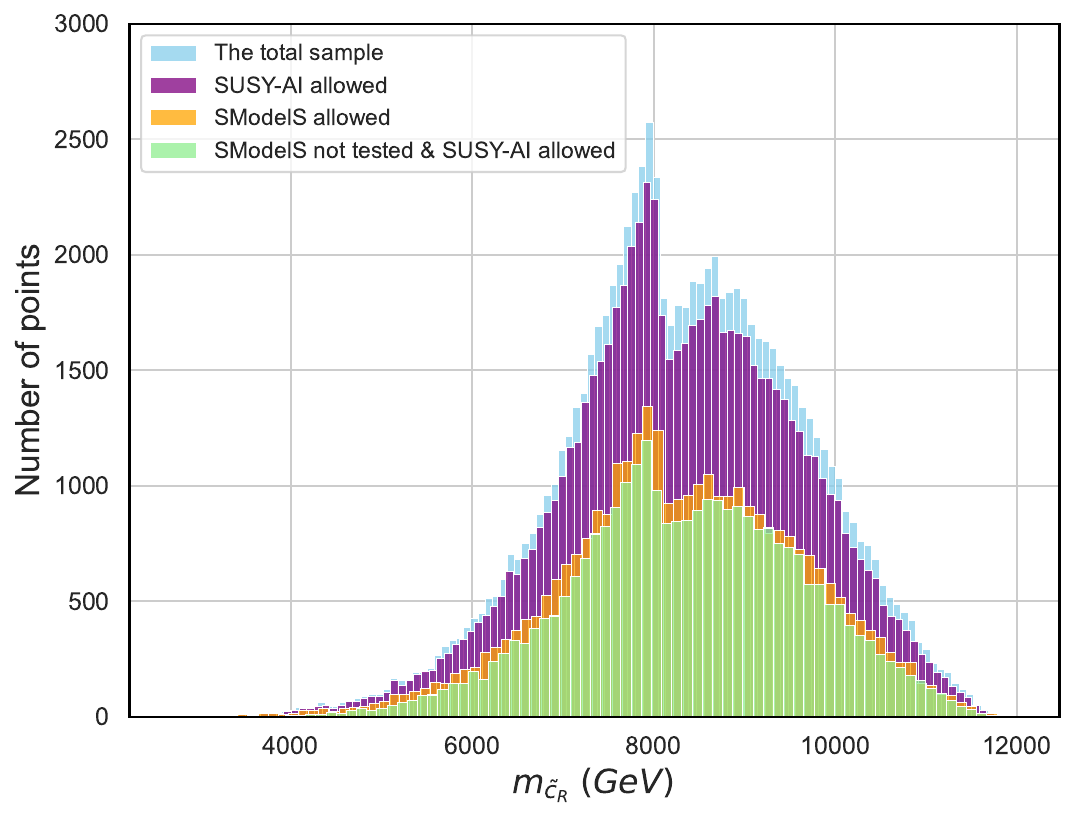} 
 \caption{The histograms of MSSM-30 predictions for observables, comparing the distributions for samples that survive SModelS and SUSY-AI constraints. From top to bottom: $d_e$, $\log(\Omega_{\rm{DM}} h^2 )$, $m_{\tilde{\chi}^0_1}$, $m_{\tilde{\chi}^+_1}$, $m_H$, $m_{\tilde{b}_1} \sim m_{\tilde{u,e,d}_R} \sim m_{\tilde{t}_1}$,  $m_{\tilde{c}_L} \sim m_{\tilde{d,s,u}_L}$ and  $m_{\tilde{c}_R} \sim m_{\tilde{s}_R} \sim m_{\tilde{\tau}_2}$.  Here the symbol $\sim$ means that the distributions are similar.} 
  \label{observs1}
\end{figure}

\begin{figure}[!ht] 
  \includegraphics[angle=0, width=0.43\textwidth]{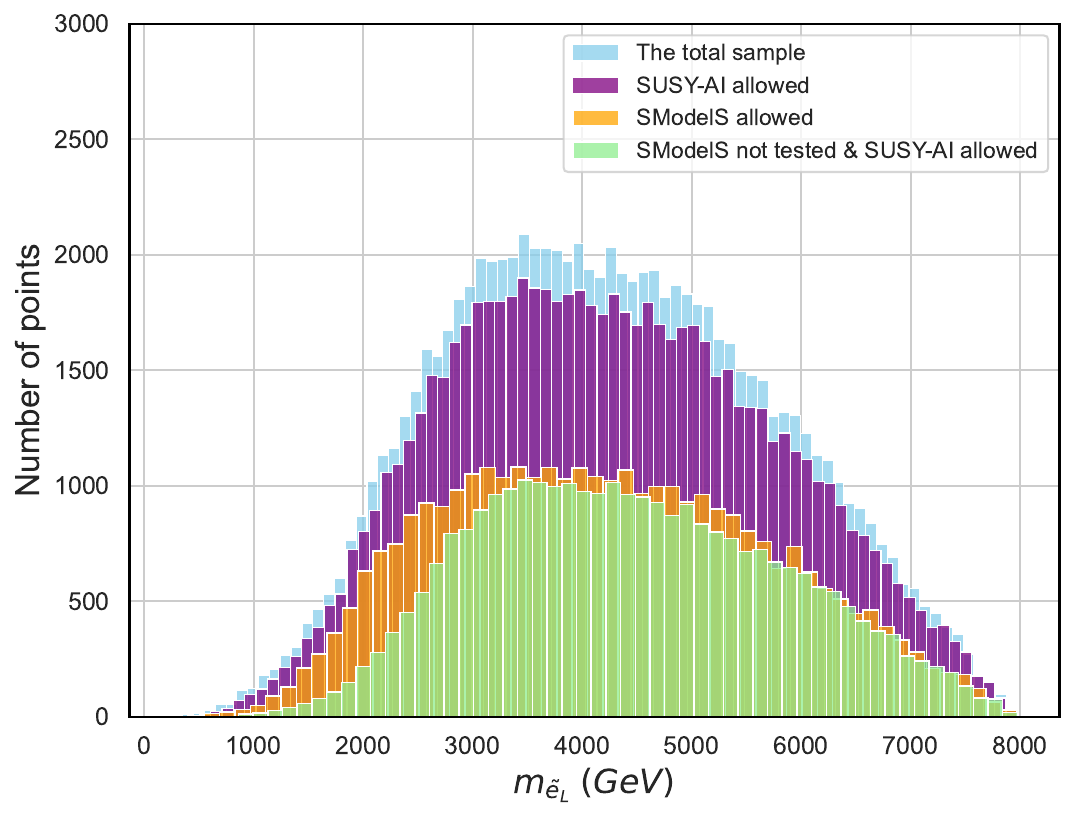} 
  \includegraphics[angle=0, width=0.43\textwidth]{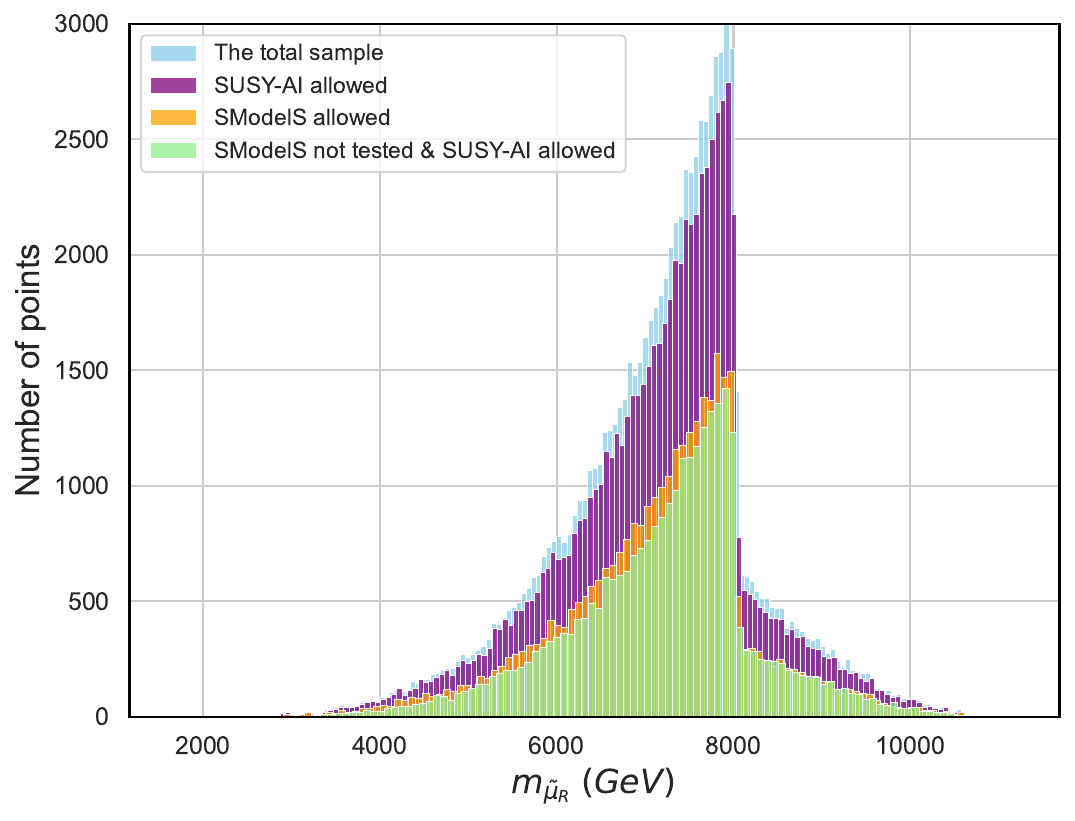} \\
  \includegraphics[angle=0, width=0.43\textwidth]{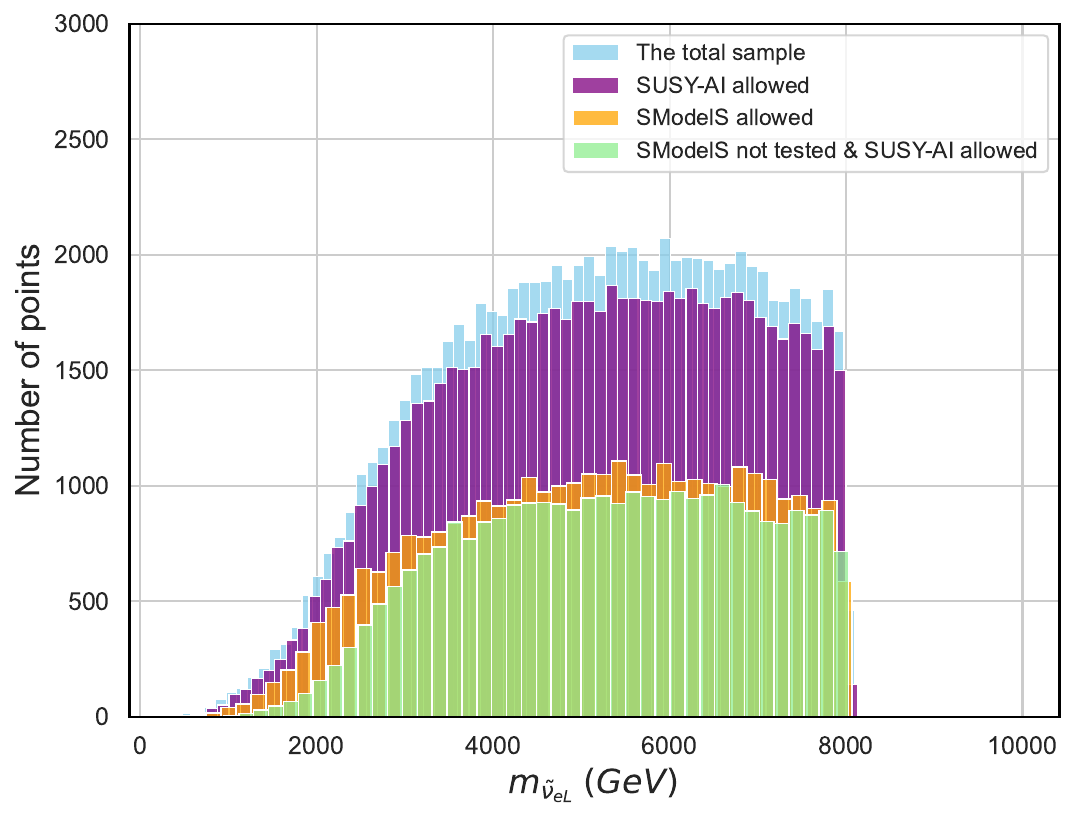}
 \includegraphics[angle=0, width=0.43\textwidth]{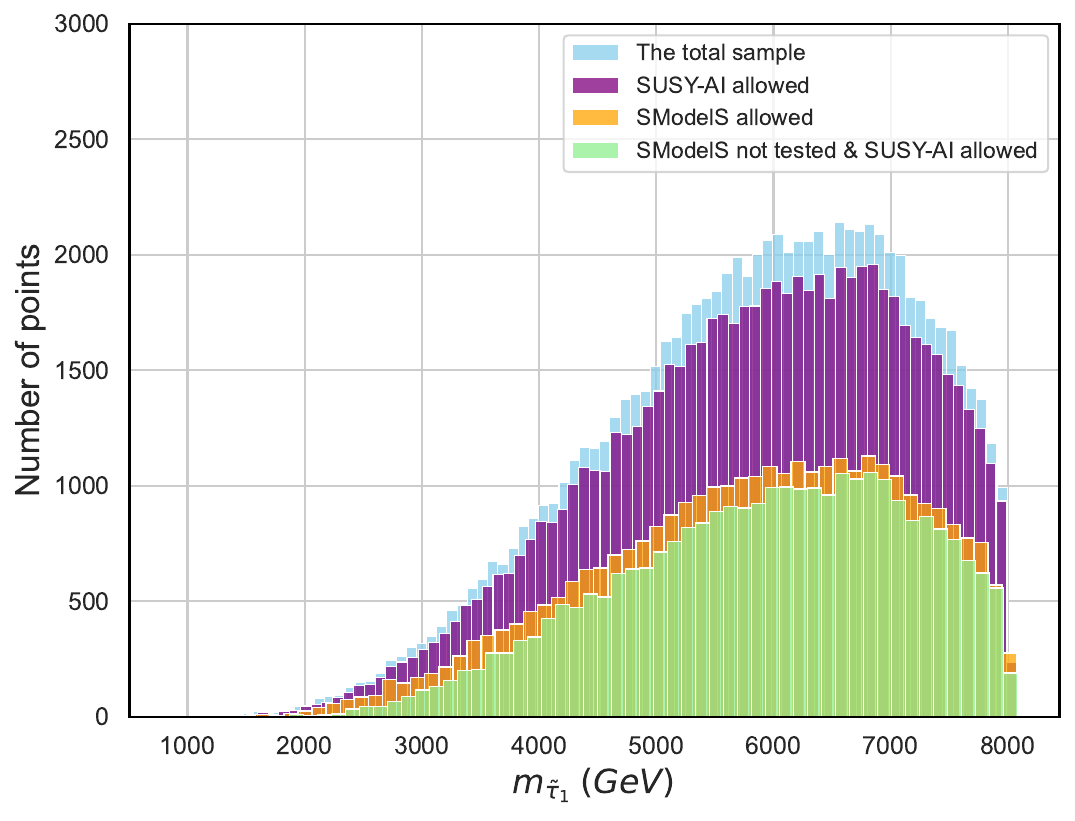} 
     \caption{The histograms of MSSM-30 predictions for observables, comparing the distributions for samples that survive SModelS and SUSY-AI constraints. From top to bottom: $m_{\tilde{e}_L}$, $m_{\tilde{\mu}_R} \sim m_{\tilde{\nu}_{\tau_L}}$, $m_{\tilde{\nu}_{eL}} \sim m_{\tilde{\nu}_{\mu}}$ and  $m_{\tilde{\tau}_{1}}$. Here the symbol $\sim$ means that the distributions are similar.} 
  \label{observs2}
\end{figure}

\paragraph{Limits from search for electron EDM}  
\begin{figure}[!t] 
\includegraphics[angle=0, width=0.475\textwidth]{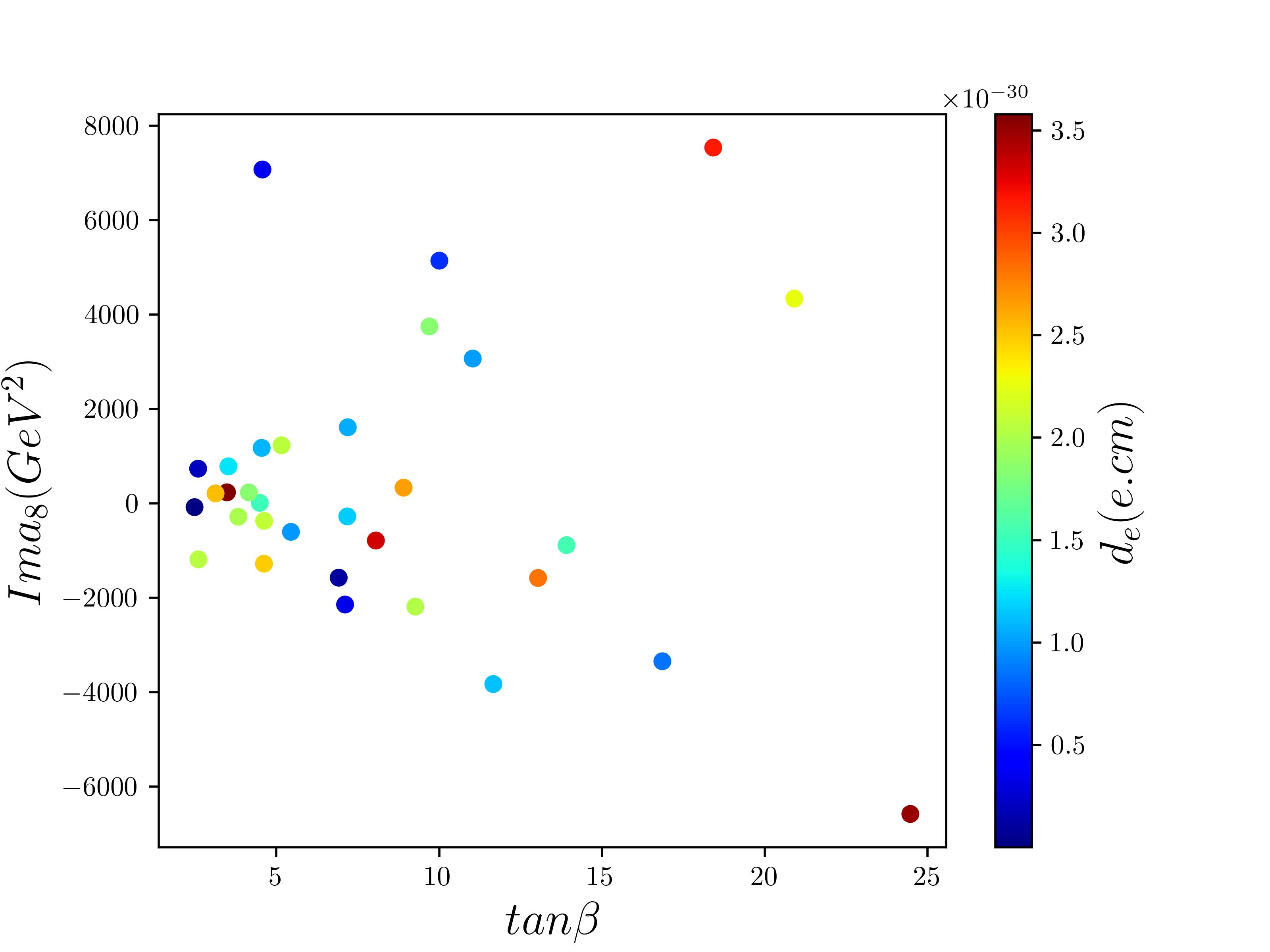}
  \includegraphics[angle=0, width=0.475\textwidth]{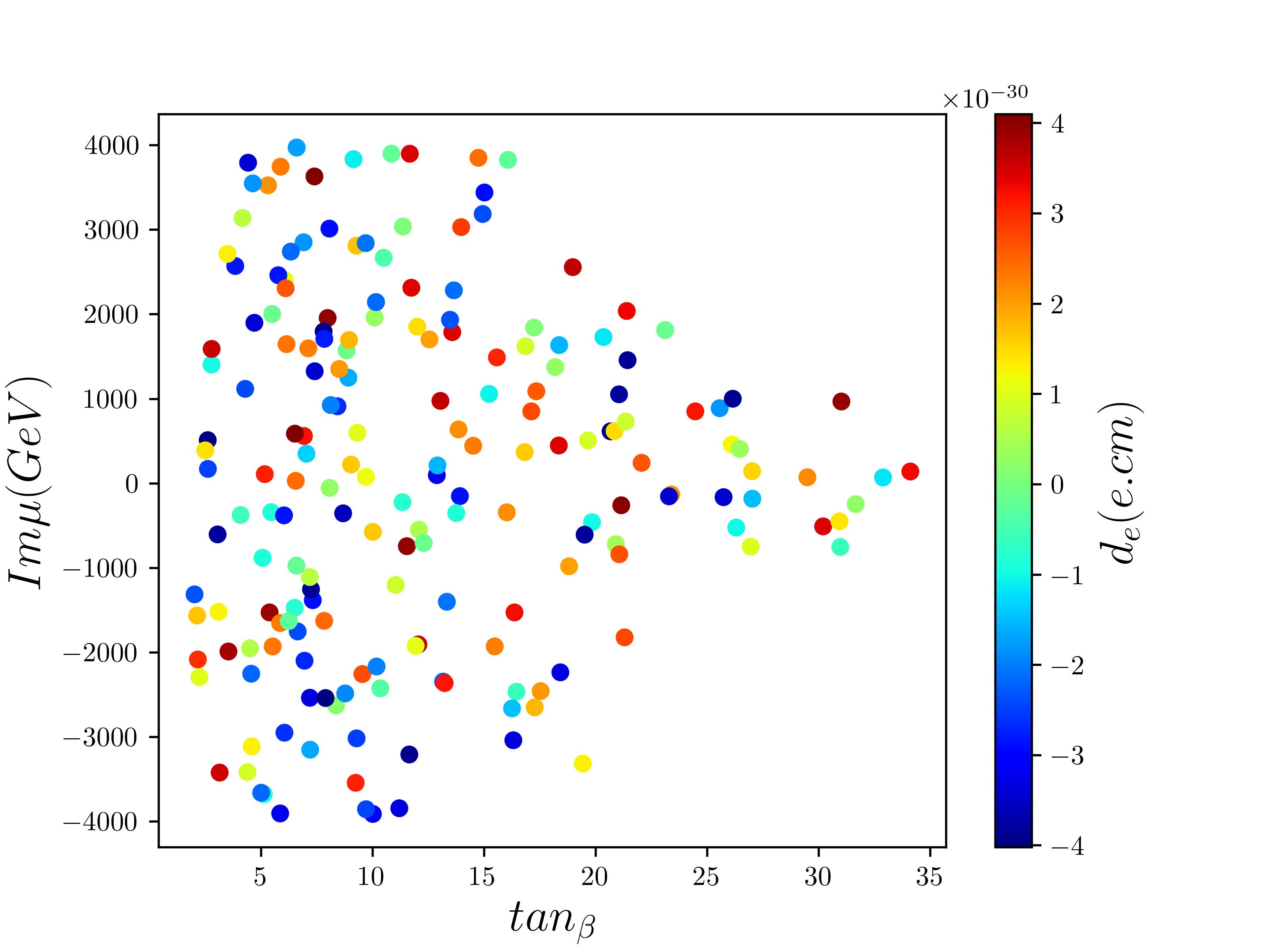}\\
  \includegraphics[angle=0, width=0.475\textwidth]{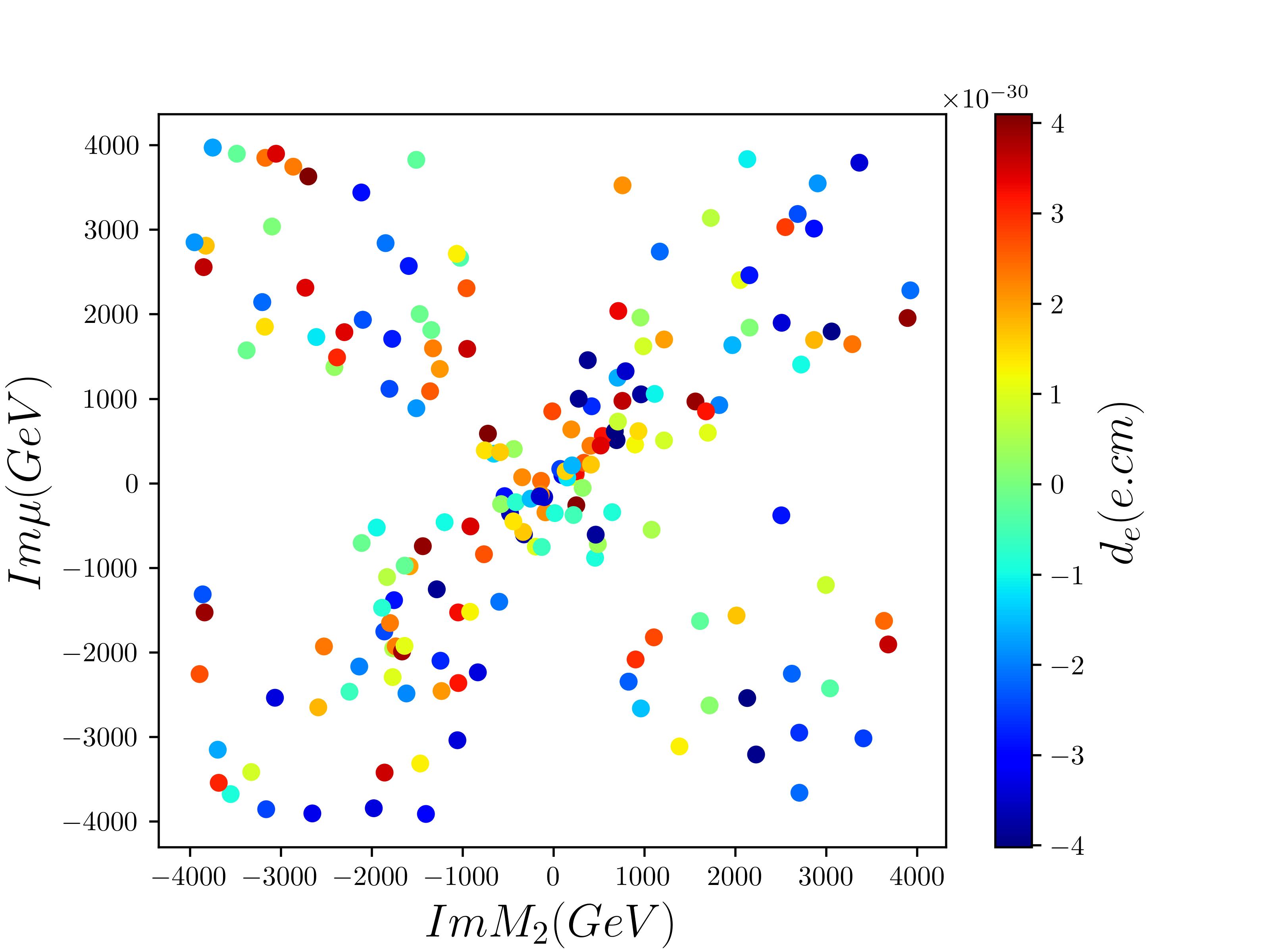}
  \includegraphics[angle=0, width=0.475\textwidth]{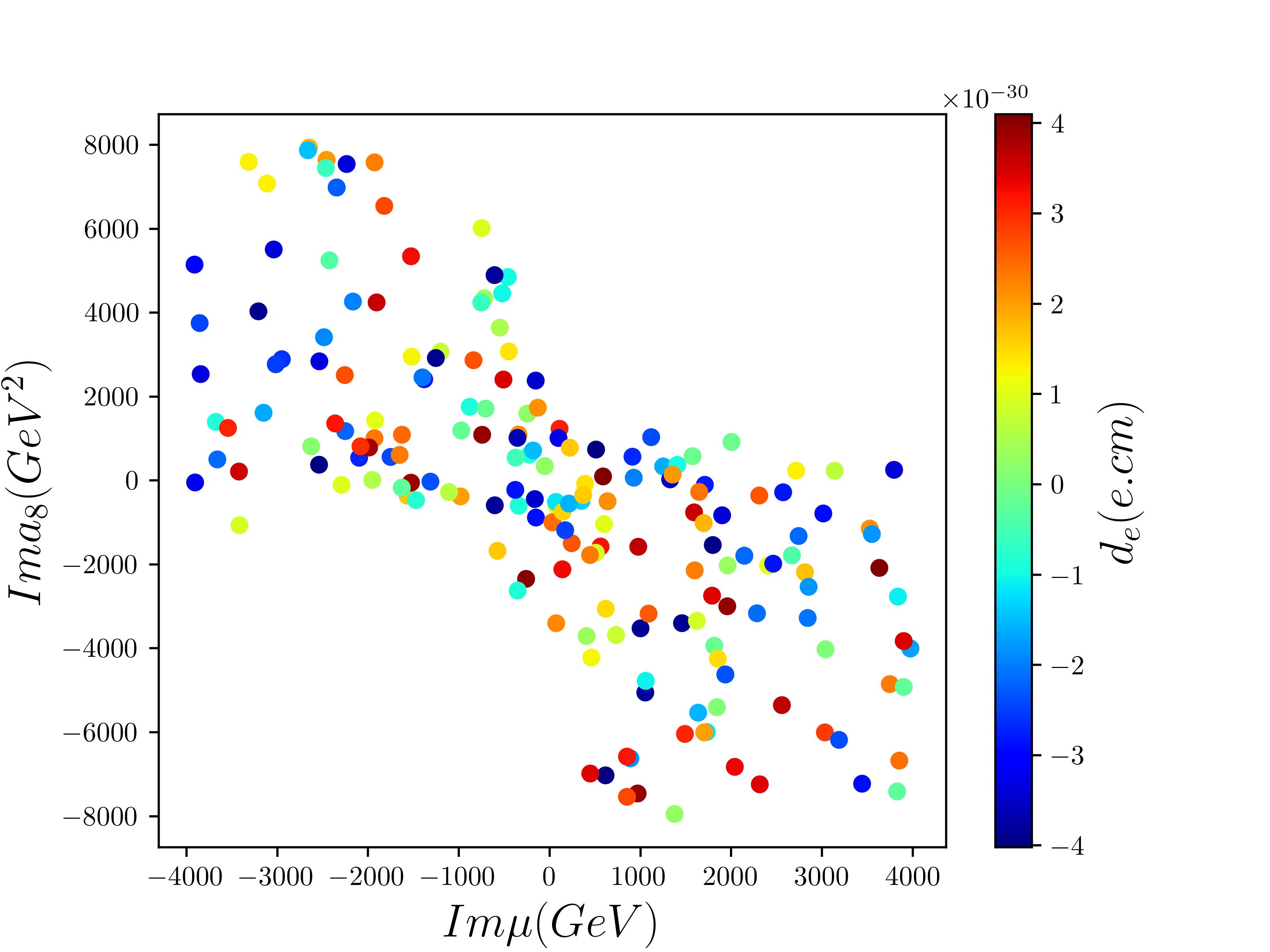}
  \caption{In the first panel of this figure we present the plane $\tan\beta$ vs $\rm{Im} a_8$ [GeV$^2$] for the case where only the imaginary trilinear term related to the electron, $a_8$, is present. In the second panel, we present the plane $\tan\beta$ vs the imaginary part of $\mu$. The third panel shows  the imaginary part of $\mu$ vs imaginary part of $M_2$ and the last part  imaginary part of $\mu$ vs imaginary part of $a_8$. From the second panel,  all the contributions in \eq{mfvpar30} are present. } 
  \label{edm_related}
\end{figure}

In \cite{Kaneta:2023wrl}, one of the authors considered EDM bounds in the light of the impressive experimental improvements in the limits on the EDMs. There, the general contributions to EDMs at one-loop and two-loops were considered and applied to some scenarios.  The contribution in the general MSSM at one-loop  from gluinos, neutralinos and charginos are given respectively in Eqs. (\ref{eq:gluinocont}, \ref{eq:neutf}), and   \eq{eq:chargf}
while the complete list of contributions at two-loops from Barr-Zee diagrams is considered in appendix B of \cite{Kaneta:2023wrl}. For completeness we have included the only two-loop diagram that is relevant for our study and is given in \eq{eq:HiggsGZW}.

In the CMSSM, it is relatively easy to understand the contributions to electron EDMs because they can be parameterised in terms of two phases, the phase of the $\mu$ term, $\theta_\mu$ and a single phase  for the trilinear terms, $\theta_A$.   This is a well defined model and can have far reaching observable consequences. In the case of the MSSM-30, many more phases affecting the EDMs come into play. 
From Eqs.~(\ref{mfvpar30}-\ref{pars}), we can see that the relevant phases are the phases for $M_1$, $M_2$, $\mu$ and the phases for the trilinear terms, by the introduction of the parameters $\tilde{a}_8$, for the $A'_E$  term and  $\tilde{a}_4$, $\tilde{a}_5$,  $y_4$ and  $y_5$ for the  $A'_U$  and $A'_D$ terms respectively.  In a more general set up an additional source for EDM phases comes from the squared mass terms $M^2_E$, $M^2_U$ and $M^2_D$, although only for the gluino contributions the phases associated to $M^2_U$ and $M^2_D$ will contribute independently from the phases of  $M_1$, $M_2$ and $\mu$  to the neutron and proton EDMs. For the electron EDM the phases on $M^2_E$ do not contribute independently from the phases of  $M_1$, $M_2$ and $\mu$ to the neutralino or chargino contributions and hence not considering the phases on $M^2_E$  will not be a non representative consideration of the phases affecting the electron EDM.

Concentrating for the moment on the EDM of the electron, we then consider the effect of the phases of $M_1$, $M_2$, $\mu$  and $\tilde{a}_8$ on the EDMs and make general comments similar to the CMSSM case in \cite{Kaneta:2023wrl}. At low $\tan\beta$ the neutralino contributions dominate the EDM. Since the contributions are proportional to $m_{\tilde{X}_i^0}/m^2_{\tilde f_k}$, where $m_{\tilde{X}_i^0}$ is the mass of the i-th neutralino and   $m^2_{\tilde f_k}$  is the mass of the k-th squark  the stronger the hierarchy between $m_{\tilde{X}_i^0}$ and $m^2_{\tilde f_k}$ the bigger the contribution, of course specially if  $m_{\tilde{X}_i^0}$ is light enough. The phase entering here is basically the phase of the the combination $N^*_{ik} S_{f_{1k}}$ where $N$ is the matrix diagonalising the neutrino mass $M_N$,\,   $N^\dagger M_{N} N*=$diag.,  $S_f$ is the matrix diagonalising the squared mass matrix (containing the A terms), $S^\dagger_f M^2_{\tilde f}S_f$.  Depending then on the hierarchy of $M_1$, $M_2$ and $\mu$ one phase will dominate over the others in $N^*_{ik}$ and for the electron EDM.

For medium ($\sim 10$) to large $\tan\beta$ the chargino contribution begins to be important and even cancels the neutralino contribution.  At large  $\tan\beta$ a Bar-Zee diagram ( two-loop) involving a photon, sfermion and chargino, that we will refer as $D_{\gamma A^0 \tilde f} $ (it is in fact a two-loop diagram involving also charginos, see Fig. 2 of \cite{Olive:2005ru} and \eq{eq:gluinocont} in the appendix),  can be sizable and even dominate the total electron EDM given the cancellation from charginos and neutralinos.  The one-loop chargino contribution depends only on the phases of $M_2$ and $\mu$, while the  diagram $D_{\gamma A^0 \tilde f}$ involves all the phases. This has the effect of potentially cancel the effect of the neutralino or the chargino contributions. 

In Fig.~(\ref{edm_related}) we present the points that satisfy the current electron EDM experimental bound. We begin with a plot of $\tan\beta$ vs $\rm{Im} a_8$, for a case that only that phase is not zero.  At low $\tan\beta$ we can see that the density of points that satisfy the experimental bound is larger than for bigger values, which it is expected since for larger values of $\rm{Im}\, a_8$ represent a bigger contribution to the electron EDMs, except when cancellations occur. We can see that these cancellations should be occurring at large $\tan\beta$, but as we expect should not be the norm (due to the different neutralino contributions and the contributions from the diagram $D_{\gamma A^0 \tilde f}$, which involve both the diagonalisation matrices of the charginos and the s-fermion masses, involving the $A$ terms) and so this should be explaining the lower density of points that in the case of lower $\tan\beta$. In the second panel, we present the  $\tan\beta$ vs $\rm{Im} \mu$, but this time all phases are not zero. As expected, the lower $\tan\beta$, the more points allowed, but this time the distribution is different since now all phases of $M_1$, $M_2$, $\mu$ are non zero and although for lower values of  $\tan\beta$ the cancellations among all no much cancellations take place. But when $\tan\beta$  grows, these cancellations become less effective as the different phases contribute quite differently when  $\tan\beta$ is large. 
In the third panel of Fig.~(\ref{edm_related}), we show $\rm{Im}\, \mu$ vs $\rm{Im}\, M_2$. The first thing to notice is that the distribution is fairly symmetrical along the four quadrants, so we can discuss only the first quadrant, that is both $\rm{Im}\, \mu$ and $\rm{Im}\, M_2$ positive. The distribution of points is such that large  values of $\rm{Im}\, M_2$  seem to be excluded for small values of $\rm{Im}\, \mu$ and vice versa, this is because an additive contribution from $\rm{Im}\, \mu$ and $\rm{Im}\, M_2$ can easily saturate the limit so the overall contribution from $\rm{Im}\, \mu$ and $\rm{Im}\, M_2$  needs to remain small. The last panel  shows the imaginary part of $a_8$ vs the imaginary part of $\mu$, where we see they have a similar behaviour as the third panel, regarding the interference between $\rm{Im}\, \mu$ and $\rm{Im}\, M_2$, except that when they have the same sign a positive interference takes place, saturating effectively the $d_e$ limit even for small phases. 

When only one phase is in play we can set bounds on the lightest sfermion, chargino and neutralino, although this is not a realistic case since once renormalisation group effects enter into consideration the phases propagate in all sectors.  However, for comparison to the CMSSM, where $\mathcal{O}(10) $ TeV masses are allowed, in our case electron EDMs does not exclude masses of $\mathcal{O}(2) $ TeV. For example, in the CMSSM with $m_{1/2}=5.4$ TeV and $m_0=12.5$ TeV (with resulting lightest neutralino and chargino around $6$ TeV) the electron EDM saturates the current experimental bound ($|d_e|<4.1\times 10^{-30}$ e cm) for a $\mu$ phase $\theta_\mu=1.4 \times 10^{-2}$ while for the MSSM-30   the lightest neutralino and chargino can be as low as $1.96$ TeV for a phase $\theta_\mu= 1.7 \times 10^{-1}$ and $|d_e|=2.0 \times 10^{-30}$.

\paragraph{Dark matter relic density and direct search limits:} 
Here the discussions are centred around dark matter-related constraints on the MSSM-30 sample. The sample points gathered already pass \texttt{micrOMEGAs} limits on the neutralino LSP and nucleons scattering cross section, $\sigma^{SI}$. Given that the MSSM30 spectra are necessarily saved in the SLHA-2 form, we had to use \texttt{MadDM}~\cite{2012.09016,1505.04190,1308.4955} also for computing the LSP dark matter properties. In Fig.~\ref{fig.sigmas}, the $m_{\tilde{\chi}^0_1}$ versus $\sigma^{SI}$ scatter plots are shown for four different cases featuring all sample points (top-left plot); \texttt{SModelS} not-tested but \texttt{SUSY-AI} passed (top-right plot); \texttt{SModelS} not-tested, \texttt{SUSY-AI} and electron EDM limit passed; and \texttt{SModelS} not tested, \texttt{SUSY-AI} plus electron EDM plus DM relic density limits passed sample points. These are displayed in contrast to the XENON1T~\cite{1708.07051}, the XENONnT~\cite{2007.08796}, next-generation XENON projection~\cite{2203.02309}, and the neutrino backgrounds~\cite{0706.3019} to DM searches contour limits. 
\begin{figure}[!t] 
  \includegraphics[angle=0, width=.450\textwidth]{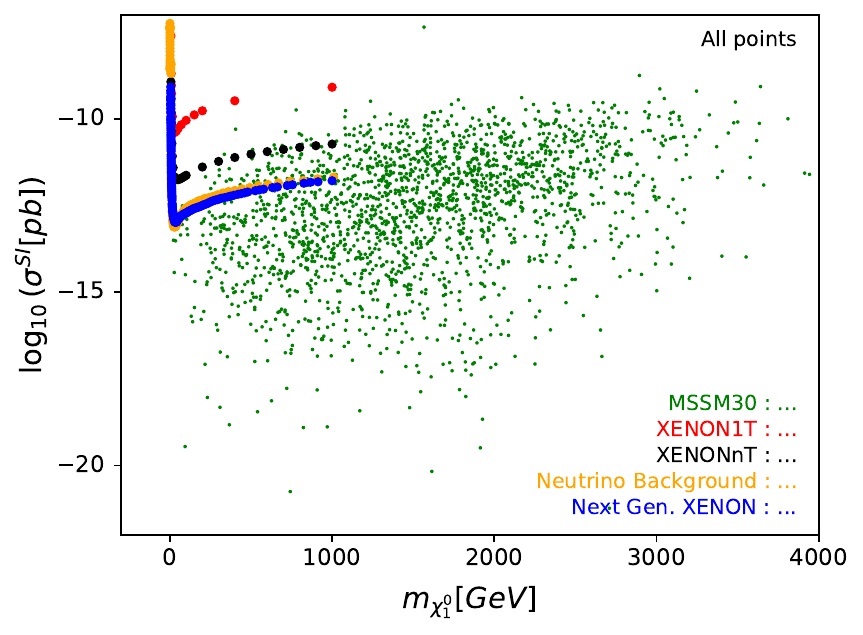}
  \includegraphics[angle=0, width=.450\textwidth]{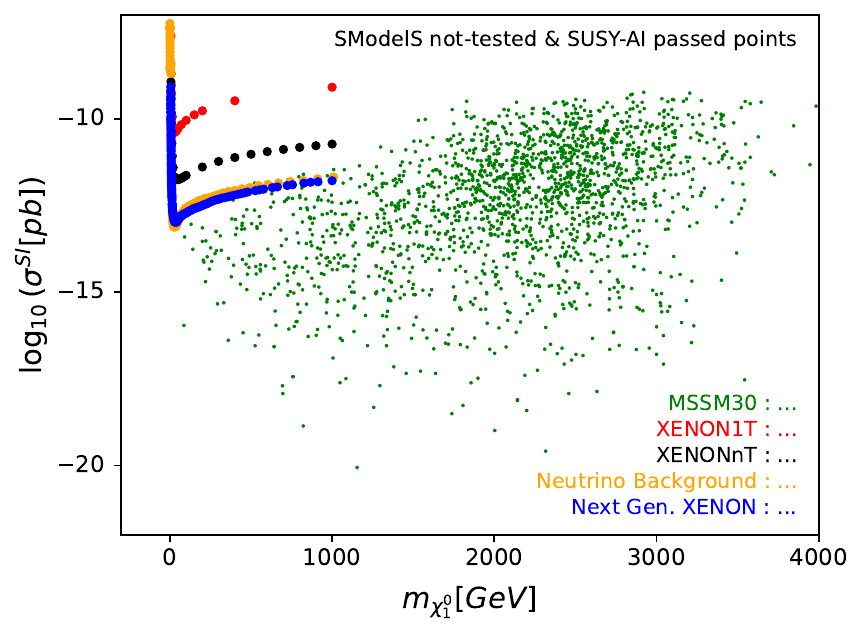}
  \includegraphics[angle=0, width=.450\textwidth]{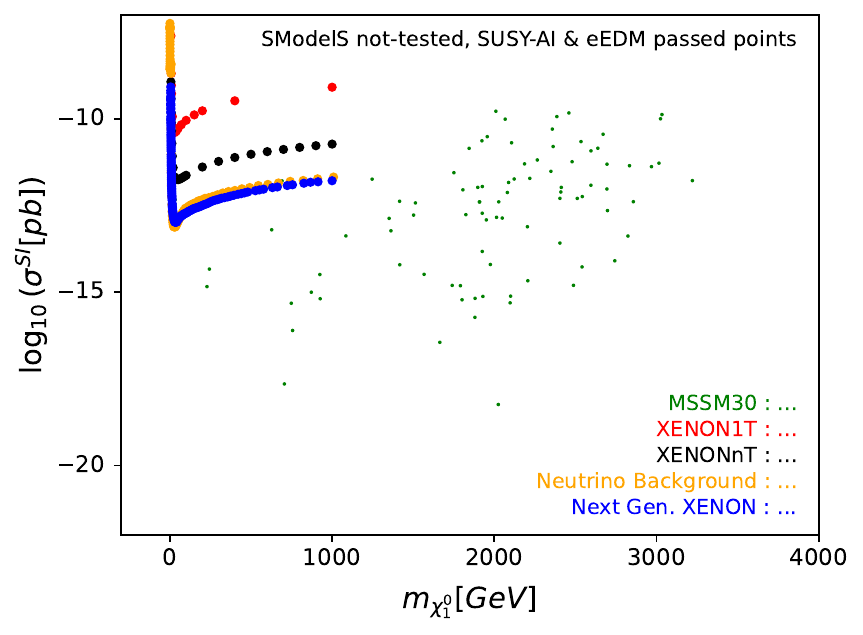}
  \includegraphics[angle=0, width=.450\textwidth]{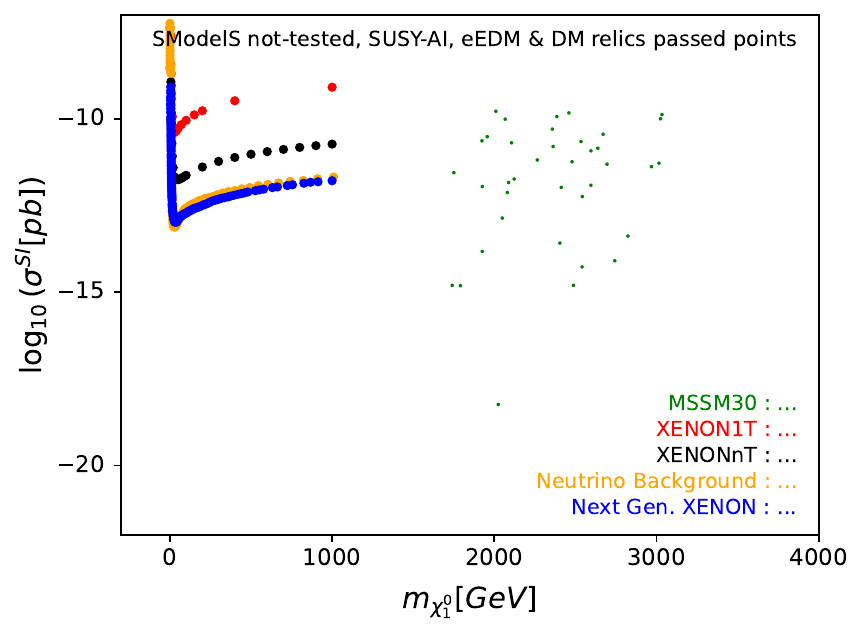}
  \caption{MSSM-30 samples neutralino mass versus the neutralino-nucleons spin-independent scattering cross-section. The dotted curves represent XENON1T, XENONnT, next-generation XENON projections, and ``Neutrino Background'' upper limits on the cross section.}  
  \label{fig.sigmas}
\end{figure}

The sample points not tested by \texttt{SModelS} but allowed by \texttt{SUSY-AI} can be considered as LHC unprobed MSSM-30 points, should we assume that all the SUSY limits are incorporated into \texttt{SModelS} and \texttt{SUSY-AI}. Requiring the electron EDM and DM relic density limits reduces further the number of points shown in the second row of the Figure. Benchmark points can be extracted in this manner, say from one of the last two samples depending on whether compliance to the relic density limit is considered necessary or not. In Tab.~\ref{all_benchmarks} we show MSSM-30 points with the possibility of charginos being long-lived and the decay (computed using \texttt{FeynHiggs}).

The chargino masses are given by $m^2_{\tilde{\chi}^{\pm}_{1(2)}} = \frac{1}{2}\left[\left(|M_2|^2 + |\mu|^2 + 2m^2_W\right) -(+) \sqrt{\Delta}\right]$ where $\Delta = \left(|M_2|^2 + |\mu|^2 + 2m^2_W\right)^2 - 4|M_2\mu - m^2_Wsin(2\beta)|^2.$ For \begin{math}|M_2|\gg m_W\end{math}, \begin{math}|\mu|\gg m_W\end{math}, and $|M_2|>|\mu|$ then $m^2_{\tilde{\chi}^{\pm}_{1}} \thickapprox |\mu|^2$ and $m^2_{\tilde{\chi}^{\pm}_{2}} \thickapprox |M_2|^2.$ The same approach of using the tree-level masses for the lightest neutralino shows that the long-lived nature of the benchmark points is related to $SU(2)$ symmetry and happen in the no-symmetry breaking limit.
    
\begin{table}[!ht]
\centering
\begin{tabular}{|l|l|l|l|l|l|}
 \hline
 \multicolumn{5}{|c|}{MSSM-30 Benchmark Points}\\
 \hline
 \multicolumn{1}{|c|}{}&
 \multicolumn{1}{|c|}{B1}&
  \multicolumn{1}{|c|}{B2}& 
 \multicolumn{1}{|c|}{B3}&  
 \multicolumn{1}{|c|}{B4}\\ 
 \hline
  $\tan \beta$ & 4.40& 4.39 & 3.13 & 3.83\\
  $m_{H^\pm}$ \footnotesize{[GeV]} & 3457.92& 4142.80 & 3764& 896\\
  $m_{\chi_1^0}$ \footnotesize{[GeV]} & 77.28& 89.21 & 1925& 1926\\
 $m_{\chi_1^{\pm}}$ \footnotesize{[GeV]}& 78.07 & 91.46 & 1953& 1951\\
 $m_{\tilde{g}}$ \footnotesize{[GeV]} &  2041.44& 3209.37 & 3590& 3563 \\
 $Real(M_{2})$ \footnotesize{[GeV]}& -63.61 & 75.87& -115 & -944\\
 $Im(M_{2})$ \footnotesize{[GeV]} & -21.29& 42.37& -1860 & -1595 \\
 $Real(\mu)$ \footnotesize{[GeV]} & 2486.92 & 1998.20 & -14 & -1756\\
 $Im(\mu)$ \footnotesize{[GeV]}& -1345.76 & 1136.39 & -3421 & 2572\\
  \hline
 $|d_{e}|$ \footnotesize{[e.cm]} $\times 10^{-30}$ & $3.53$ & $2.54$ & $3.53$ & $2.83$\\
  \hline 
\end{tabular}
\caption{MSSM-30 points which passed all experimental constraints described (with \texttt{micrOMEGAs} dark matter relic density) with small $\Delta m = m_{\chi_1^{\pm}} - m_{\chi_1^0}$ for the first two benchmarks. All squarks are heavy and well into the TeV/multi-TeV scales.\label{all_benchmarks}}
\end{table}
The Higgs production cross sections and decay rates for these benchmark points and all the other sample points saved as root files can be accessed at~\cite{DVN/JFTNCM_2023}. The cross sections for the MSSM-30 Higgs productions at 13 TeV LHC are computed using \texttt{SusHiMi-1.7.0} and \texttt{FeynHiggs} with the $PDF4LHC15\_nlo\_mc$ and $PDF4LHC15\_nnlo\_mc$~\cite{Butterworth:2015oua} parton distribution function sets.

\section{Summary}
We have used the 30-parameters R-parity conserving MSSM phenomenological framework called MSSM-30 in developing a computing pipeline for extracting benchmark spectra for further collider studies. The model is constructed with minimal flavour violation hypothesis implemented in a systematic way, incorporating flavour and CP-violation parameters, in order to compare with results from experiments. The MSSM-30 sample of spectra were generated via random scans in parameter space. As a proof of principle, benchmark points with long-lived charginos (as in less than a GeV mass difference to neutralino LSP) and lightest CP-even Higgs decays to BSM particle possibilities are extracted. These are compatible with limits from the LHC, low-energy constraints from electroweak physics, B-physics, dipole moment of the electron and cold dark matter relic density measurements. 
Table \ref{all_benchmarks} exemplifies the diversity of spectra that our search was able to identify. Intriguing spectra  with the lightest neutralinos lighter than the W-boson mass, while the rest of the spectra around 3 TeV, deserve dedicated studies.

The lessons are clear regarding the systematic search for supersymmetric scenarios that are still accessible to be probed at the LHC. 
The search for supersymmetric scenarios cannot be given up at this time, the current developments in machine-learning tools and well motivated ways to systematically reduce the number of parameters, like the MSSM-30 presented here, have the power to uncover parameter space regions that serve as a justification of the validity of a light supersymmetric scenario. In the future, we hope to couple this phenomenological framework with machine-learning tools for a tailored exploration of specific BSM features of interest for LHC and future collider searches for BSM particles. 

\paragraph{Acknowledgement}: Thanks to CERN-TH for access to the HTCondor computing resource. 

\appendix 

\section{SUSY Notation}

We follow the notation of \cite{Kaneta:2023wrl}, that for completeness we include here. The matrices diagonalising the sfermions are labelled by $S$, 
\bea
\label{eq:Sq}
S_f^\dagger M^2_{\tilde f}S_f={\rm{diag.}}\left(m^2_{\tilde f_1},\, m^2_{\tilde f_2}\right),
\eea
for $m^2_{\tilde f_1} < m^2_{\tilde f_2}$. For the neutralinos, we use the convention $\ 
{\mathcal{L}}=-\frac{1}{2}\widetilde\psi^0
{\mathcal{M}}_{\widetilde\psi^0} (\widetilde \psi^0)^T\ $ 
$+\ \quad {\mathrm{h.c.}} \;$,
where the gauge eigenstates are 
$\widetilde\psi^0 = (-i\tilde b, -i\widetilde w, \tilde h_d^0, \tilde h_u^0)$ and the mass matrix is given by
\begin{equation}
\mathcal{M}_{\widetilde\psi^0} = \left[
\begin{array}{cccc} 
M_1 & 0 &
-M_Z\ \cbeta\ \sw & M_Z\ \sbeta\ \sw \\ 0 & M_2 & M_Z\ \cbeta\ \cw &
-M_Z\ \sbeta \ \cw \\ -M_Z\ \cbeta\ \sw & M_Z\ \cbeta \ \cw & 0 & -\mu \\
M_Z\ \sbeta\ \sw & -M_Z\ \sbeta\ \cw & -\mu & 0
\end{array} \right],
\label{eq:mchi0}
\end{equation}
where the diagonal matrix is given by $M_{\chi^0}=N^\dagger {\cal M}_{\widetilde\psi^0} N^*$,
and the mass eigenstates, $\tilde\chi^0_i$, are given by $\tilde \chi^0_i=(\tilde\psi^0 N)_{i}$. For charginos, the notation is that of \cite{Haber:1984rc}  with the following identification 
$\tilde\psi^{\pm}=\left(-i \tilde \omega^\pm, \tilde h^\pm_u \right)$, with the Lagrangian $\mathcal{L} =
  -\frac{1}{2}\left[ \widetilde \psi^+ V V^+
 {\cal{M}}_{\widetilde \psi^+} U^* U^T (\widetilde \psi^-)^T + {\rm{h.c.}}\right]$ (where $V$ and $U$ are unitary matrices) and the mass matrix
\begin{equation}
\label{eq:int_basis_charg}
{\cal M}_{\widetilde\psi^+} = \left( \begin{array}{cc} M_2 &
\sqrt2\,M_W\, \cbeta\\\sqrt2\,M_W\, \sbeta & \mu\end{array}\right)\, ,
\end{equation}
which is hence diagonalised by $U$ and $V$, according to  $M_{\rm{diag.}}=M_{\tilde \chi^{\pm}} = V^+ {\cal M}_{\widetilde\psi^+} U^*$.
For the diagonalisation of the CP even mass Eigenstates we choose
\bea
{\mathcal{H}}=\left(
\begin{array}{c}
h_1 \\
h_2
\end{array}
\right)=
Z_H \,
\left(
\begin{array}{c}
h \\
H
\end{array}
\right),
\eea
for $h$ and $H$ being respectively the CP neutral mass Eigenstates.  In this notation then, the gluino contribution to the quark EDMs, denoted by $d_q^{\tilde g}$, is given by
\begin{align}
\label{eq:gluinocont}
    \frac{d_q^{\tilde g}}{e} &=
    -\frac{2\alpha_s}{3\pi}\sum_{k=1}^2{\rm Im}[S^*_{q2k}S_{q1k}]\frac{m_{\tilde g}}{m^2_{\tilde{q}_k}}Q_{\tilde q}B(m^2_{\tilde g}/m^2_{\tilde{q}_k}),
\end{align}
where $Q_f$ is the electromagnetic charge of fermion or sfermion, $f$. The neutralino contribution to the fermion EDMs, $d^{\widetilde{\chi}^0}_f$, is given by
\begin{align}
\label{eq:neutf}
    \frac{d^{\widetilde{\chi}^0}_f}{e} &=
    \frac{\alpha_{\rm em}}{4\pi\sin^2\theta_W}\sum_{k=1}^2\sum_{i=1}^4{\rm Im}[\eta_{fik}]\frac{m_{\tilde{\chi}^0_i}}{m^2_{\tilde{f}_k}}Q_{\tilde f}B(m^2_{\tilde{\chi}^0_i}/m^2_{\tilde{f}_k}),
\end{align}
where $\alpha_{\rm em}\equiv e^2/4\pi$, and
\begin{align}
    \eta_{fik} &=
    \left[-\sqrt{2}\left\{\tan\theta_W(Q_f-T_{3f})N^*_{1i}+T_{3f}N^*_{2i}\right\}S_{f1k}+\kappa_f N^*_{bi}S_{f2k}\right]\nonumber\\
    &\times\left(\sqrt{2}\tan\theta_W Q_f N^*_{1i}S^*_{f2k}-\kappa_f N^*_{bi}S^*_{f1k}\right), \nonumber\\
    \kappa_u &= \frac{m_u}{\sqrt{2}M_W\sin\beta},\;\;\;
    \kappa_{d,e} = \frac{m_{d,e}}{\sqrt{2}M_W\cos\beta},
\end{align}
with $b$ referring to the Higgsino contribution where $b=3$ for $f=d, e$ with $T_{3f}=-1/2$ and $b=4$ for $f=u$ with $T_{3f}=+1/2$. The function $A(r)$ is given in \eq{fct:Br}.  The chargino contribution to the $u$, $d$, and electron EDMs is as follows:
\begin{align}
\label{eq:chargf}
    \frac{d^{\tilde\chi^+}_u}{e} &=
    -\frac{\alpha_{\rm em}}{4\pi\sin^2\theta_W}\sum_{k=1}^2\sum_{i=1}^2{\rm Im}[\Gamma_{uik}]\frac{m_{\tilde{\chi}^+_i}}{m^2_{\tilde{d}_k}}\left[Q_{\tilde d}B(m^2_{\tilde{\chi}^+_i}/m^2_{\tilde{d}_k})+(Q_u-Q_{\tilde d})A(m^2_{\tilde{\chi}^+_i}/m^2_{\tilde{d}_k})\right],\\
    \frac{d^{\tilde\chi^+}_d}{e} &=
    -\frac{\alpha_{\rm em}}{4\pi\sin^2\theta_W}\sum_{k=1}^2\sum_{i=1}^2{\rm Im}[\Gamma_{dik}]\frac{m_{\tilde{\chi}^+_i}}{m^2_{\tilde{u}_k}}\left[Q_{\tilde u}B(m^2_{\tilde{\chi}^+_i}/m^2_{\tilde{u}_k})+(Q_d-Q_{\tilde u})A(m^2_{\tilde{\chi}^+_i}/m^2_{\tilde{u}_k})\right],\\
    \frac{d^{\tilde\chi^+}_e}{e} &=
    \frac{\alpha_{\rm em}}{4\pi\sin^2\theta_W}\frac{\kappa_e}{m^2_{\tilde{\nu}_e}}\sum_{i=1}^2 {\rm Im}[U^*_{2i}V_{1i}]m_{\tilde{\chi}^+_i}A(m^2_{\tilde{\chi}^+_i}/m^2_{\tilde{\nu}_e}),
\end{align}
where the function $B(r)$ is given in \eq{fct:Br} and the pre-factors are
\begin{align}
    \Gamma_{uik} &=
    \kappa_u V_{2i}S^*_{d1k}(U^*_{1i}S_{d1k}-\kappa_d U^*_{2i}S_{d2k}),\\
    \Gamma_{dik}&=
    \kappa_d U^*_{2i}S^*_{u1k}(V_{1i}S_{u1k}-\kappa_u V_{2i}S_{u2k}).
\end{align}

\begin{align}
    A(r) &=
    \frac{1}{2(1-r)^2}\left(3-r+\frac{2\ln r}{1-r}\right),\label{fct:Ar}\\
    B(r) &=
    \frac{1}{2(1-r)^2}\left(1+r+\frac{2r\ln r}{1-r}\right),\label{fct:Br}\\
    C(r) &=
    \frac{1}{6(1-r)^2}\left(10r-26+\frac{2r\ln r}{1-r}-\frac{18\ln r}{1-r}\right).\label{fct:Cr}
\end{align}

\begin{eqnarray} 
  \label{eq:HiggsGZW}   
  d_f^{\, \gamma A^0 \tilde{f}} &=&   
\frac{e\, Q_f  \alphem\, N_c}{32\pi^3}\, \frac{\tan\beta\ m_f}{m^2_{A^0}}\ \sum_{f' = t,b,\tau}\ 
c_{\tilde{f'} A^0}\, Q^2_{f'}\,\left[\, F\left(r_{\tilde{f'}_1 A^0}\right)\ -\ 
F\left(r_{\tilde{f'}_2 A^0}\right)\, \right]\, , \nonumber\\
& &  r_{\tilde{f'}_i A^0}=m^2_{\tilde{f'}_i}/{m^2_{A^0}},\
N_c, \ {\small{{\rm is\ the\  color}\ {\rm factor}}}\,,
\end{eqnarray}
where the function $F$ is defined as 
\bea
F(r_1)=\int_0^1\, dx \frac{x(1-x)}{r_1-x(1-x)}\, \log\left[\frac{x (1-x)}{r_1} \right].
\eea

\end{document}